\documentclass[fleqn,reqno]{amsart}
\usepackage[foot]{amsaddr} % puts the addresses on the first page - option 'foot' puts them into footnotes
\usepackage{comment}
\usepackage{dirtytalk}
\usepackage{mathtools}
\usepackage{bbm}
\usepackage{dsfont}
\usepackage{relsize}
\usepackage{exscale}
\usepackage[version=4]{mhchem}
\usepackage{adjustbox}

\usepackage{tikz}

\usetikzlibrary{calc,arrows,intersections}
\usetikzlibrary{patterns}

\newcommand{\norm}[1]{\lVert{#1}\rVert}

\allowdisplaybreaks

\newtheorem{innerassumption}{Assumption}
\newenvironment{assumption}[1]
  {\renewcommand\theinnerassumption{#1}\innerassumption}
  {\endinnerassumption}

\makeatletter

\renewcommand\subsubsection{\@secnumfont}{\bfseries}%
\renewcommand\subsubsection{\@startsection{subsubsection}{3}
  \z@{.5\linespacing\@plus.7\linespacing}{-.5em}%
  {\normalfont\bfseries}}
  
  \makeatother
  
\makeatletter
\def\namedlabel#1#2{\begingroup
    #2%
    \def\@currentlabel{#2}%
    \phantomsection\label{#1}\endgroup
}
\makeatother

\usepackage[hidelinks, colorlinks=true, linkcolor=blue, citecolor=red]{hyperref}
\usepackage{bm}
\usepackage{amsfonts,amssymb,amsthm,amsmath}
\usepackage{epsfig}
\usepackage{graphicx}
\usepackage{mathrsfs}
\usepackage{color} 
\usepackage[margin=1in]{geometry}
\usepackage{cases}
\theoremstyle{plain}
\newtheorem{theorem}{Theorem}
\newtheorem{proposition}[theorem]{Proposition}
\newtheorem{lemma}[theorem]{Lemma}
\newtheorem{corollary}[theorem]{Corollary}
\newtheorem*{remark}{Remark}
\newtheorem{example}{Example}

\theoremstyle{definition}

\makeatletter
\newtheorem*{rep@theorem}{\rep@title}
\newcommand{\newreptheorem}[2]{%
\newenvironment{rep#1}[1]{%
 \def\rep@title{#2 \ref{##1}}%
 \begin{rep@theorem}}%
 {\end{rep@theorem}}}
\makeatother

\newreptheorem{theorem}{Theorem}
\newreptheorem{lemma}{Lemma}

\numberwithin{theorem}{section}
\numberwithin{equation}{section}

\newcommand{\R}{\mathbb{R}}

\newcommand{\D}{{\mathcal D}}

\newcommand{\cH}{{\mathcal{H}}}
\newcommand{\nc}{\newcommand}
\nc{\e}{\epsilon}
\nc{\al}{\alpha}
\nc{\be}{\beta}
\nc{\del}{\delta}
\nc{\G}{\Gamma}
\nc{\g}{\gamma}
\nc{\ka}{\kappa}
\nc{\lam}{\lambda}
\nc{\Lam}{\Lambda}
\nc{\Om}{\Omega}
\nc{\om}{\omega} %\nc{\Omt}{\tilde{\Omega}}
\nc{\ta}{\tau}
\nc{\w}{\omega}
\nc{\io}{\iota}
\nc{\h}{\theta}
\nc{\z}{\zeta}
\nc{\s}{\sigma}
\nc{\sig}{\sigma}
\nc{\Si}{\Sigma}
\nc\vphi{{\varphi}}
\nc\lb{\lambda}
\nc\eps{\epsilon}

\nc\cchi{{\check \chi}}

\nc{\1}{{\bf 1}}

\newcommand{\Ran}{{\rm{Ran\, }}}

\newcommand{\supp}{{\rm{supp\, }}}

\newcommand{\nrm}[1]{\left|#1\right|}

\renewcommand{\part}{{\rm part}}
\nc\matter{{\rm matter}}

\nc\hc{{\rm h.c.}}
\newcommand{\DETAILS}[1]{}

\makeatletter
\ifcsname phantomsection\endcsname
    \newcommand*{\qrr@gobblenexttocentry}[5]{}
\else
    \newcommand*{\qrr@gobblenexttocentry}[4]{}
\fi
\newcommand*{\addsubsection}{%
    \addtocontents{toc}{\protect\qrr@gobblenexttocentry}%
    \subsection}
\makeatother

\usepackage[style=nature,sorting=nty]{biblatex}
\begin{document}

\title[Global Spectral Gap in Bosonic Molecular Hamiltonians]{Global Spectral Gap in Bosonic Molecular Hamiltonians}

\author[S. Gherghe]{Sebastian Gherghe${}^1$}
\email{\href{mailto:sebastian.gherghe@mail.utoronto.ca}{sebastian.gherghe@mail.utoronto.ca}}
\address{${}^1$University of Toronto, 40 St. George St., Toronto, Ontario, Canada.}
%\date{Nov 11 2020}

\begin{abstract}
    We consider a neutral bosonic molecule in the Born-Oppenheimer approximation without spin and assume the physically obvious assertion that a neutral molecule prefers to break into smaller neutral clusters. We prove the existence of a global in space/uniform spectral gap between the ground state and first excited state energies. To do so, we improve upon previous results using a different tool, the time-independent Feshbach-Schur map.
\end{abstract}

\maketitle

\tableofcontents

%%%%%%%%%%%%%%%%%%%%%%%%%%%%%%%%%%%%%%%%%%%%%%%%%%%%%%%%%%%%%%%%%%%%%%%%%%%%%%%

\section{Introduction} \label{Introduction_Section}

In this paper, we consider a neutral molecule in the Born-Oppenheimer approximation and investigate the presence of a global in space/uniform spectral gap between the ground state energy and the rest of the spectrum. Consider a molecule consisting of $N$ spinless electrons of mass $m_e$ and charges $-e$ and $M$ nuclei of masses $m_j$ and charges $+eZ_j$ for $j=1, \dots, M$. Its Schr{\"o}dinger operator is given by 
\begin{equation} \label{Hbo(y)_withCoulomb}
    H_{N,M}(y) = - \sum_{j=1}^N \frac{1}{2m_e} \Delta_{x_j} + V_e(x) + V_{en}(x,y) + V_n(y),
\end{equation}
acting on either $L^2(\mathbb{R}^{3N})$ or a symmetry subspace, and
\begin{align}
    V_e(x) &= \sum_{1 \leqslant i < j \leqslant N} \frac{e^2}{\lvert{x_i - x_j}\rvert}, \\
    V_{en}(y) &= \sum_{i=1}^N \sum_{j=1}^M \frac{e^2 Z_j}{\lvert{x_i - y_j}\rvert}, \\
    V_{n}(y) &= \sum_{1 \leqslant i < j \leqslant M} \frac{e^2 Z_i Z_j}{\lvert{y_i - y_j}\rvert},
\end{align}
are the Coulomb interaction potentials of electron-electron, electron-nuclei, and nuclei-nuclei, respectively. The operator $H_{N,M}(y)$ is parameterized by $y = (y_1, \dots, y_M) \in \mathbb{R}^{3M}$ with $y_i \neq y_j$ for $i\neq j$, the positions of the nuclei, and the positions of electrons are given by the $x = (x_1, \dots, x_N) \in \R^{3N}$ coordinates. We denote by $Z = \sum_{j=1}^M Z_j$ the total nuclear charge, and for a neutral molecule, we have
\begin{equation}
   N = Z. \label{neutralmolecule}
\end{equation}
Let $E_0(y)$ denote the ground state energy and $E_1(y)$ the infimum of the rest of the spectrum, 
\begin{align}
    E_0(y) &= \inf_{\substack{\norm{\phi} = 1 \\ \phi \in \D(H_{N,M}(y))}} \langle \phi, H_{N,M}(y) \phi \rangle, \label{E0_def_intro}\\
    E_1(y) &= \inf \left( \sigma(H_{N,M}(y)) \setminus \{E_0(y)\} \right).
\end{align}
We are interested for which $\mathcal{A} \subseteq \R^{3M}$ the uniform \textit{spectral gap condition} holds, 
\begin{equation} \label{gapcondition}
    E_1(y) - E_0(y) \geqslant \delta > 0, \quad \delta > 0 \text{ independent of }y \in \mathcal{A}.
\end{equation}

By the HVZ Theorem, the existence of such a spectral gap implies that $E_0(y)$ is an isolated eigenvalue, below the essential spectrum of $H_{N,M}(y)$. In particular, this indicates that the system is stable with respect to breaking up into smaller clusters, known as \textit{stability with respect to break up}.

We briefly discuss the challenges posed by problem \eqref{gapcondition}. For bosonic systems or systems without particle statistics, it is well-known that $E_0(y)$ is non-degenerate (see for example \cite[Theorem XIII.46]{RSv4}), ensuring the existence of a spectral gap \eqref{gapcondition} on bounded open subsets of $\R^{3M}$. However, this does not exclude the possibility of the gap closing as $\nrm{y} \to \infty$. Consequently, we must consider the system’s break-up limits into configurations of smaller clusters, which may be neutral or ionic. This is the first point of difficulty and will involve precise notation and set-up. The physically intuitive assertion that neutral systems prefer neutral clusters upon separation remains an open conjecture in mathematical quantum mechanics (see \cite{IA-IMS} and references therein), which we adopt here as an assumption.

\begin{assumption}{[LN]} \label{localneutrality_assumption_introversion}
(Local neutrality.) A neutral molecule that breaks up into groups of smaller molecules which recede from each other does so in a manner that keeps each smaller system neutral. 
\end{assumption}

To analyze the gap in such limits, we must determine the asymptotics of both $E_0(y)$ and $E_1(y)$ as $\nrm{y} \to \infty$. While the asymptotics for the ground state energy $E_0(y)$ are relatively straightforward, those for $E_1(y)$ require consideration of both the first excited states of neutral clusters and the ground state energies of ions. This adds complexity, as the space of candidate states is large and must be examined carefully.

The \textit{Local neutrality} conjecture is partly motivated by the van der Waals force, a quantum effect that arises between two neutral molecules at a certain distance\footnote{Quantum correlations between intermolecular forces and spins induce an attractive force that decays as the inverse sixth power of the distance between the molecules.}. The existence of this force was rigorously proven in \cite{MS, LT}, with its exact form under additional non-degeneracy conditions derived in \cite{IA-IMS}. The van der Waals force implies that all neutral molecules can bind in the Born-Oppenheimer approximation, meaning the ground state energy has a global minimum with respect to nuclear positions. For any molecular split, the energy remains higher than the lowest possible energy due to this force. While the minimizer is non-unique (due to translational symmetry), these minima, corresponding to molecular shapes (see for example \cite{LT}), are fundamental in quantum chemistry .

Our main theorem can be summarized as follows.

\begin{theorem}\label{maintheorem_gap_intro}
For a neutral molecule of interacting bosons with two or more nuclei ($M \geqslant 2$) and assuming Assumption \ref{localneutrality_assumption_introversion} holds, the gap condition (\ref{gapcondition}) holds with $\mathcal{A} = \R^{3M}$.  
\end{theorem}

\begin{remark} 
While electrons are fermions, nuclei can be either fermions or bosons\footnote{Nuclei of different charges are different species, in the sense that they do not obey any particle statistics upon exchange.}. However, nuclear statistics are irrelevant in this context. 
\end{remark}

\begin{remark} 
For fully fermionic systems, the ground state energy is generally degenerate, breaking the spectral gap. Our main result applies to molecules with fermionic electrons if non-degeneracy of the ground state energy is assumed for both the system and any subsystem. In such cases, the proof follows similarly, with antisymmetric wavefunctions in place of symmetric ones. Since this assumption is strong and rarely satisfied, even in simple cases like hydrogen gas \ce{H2}, we exclude it from the main theorem. 
\end{remark}

Without a non-degeneracy condition, a global spectral gap for fermionic systems is unlikely, as these systems typically exhibit eigenvalue crossings, where eigenvalues of $H_{N,M}(y)$ may intersect or merge into the continuous spectrum. Avoided crossings, where eigenvalues come very close but remain distinct, also challenge the adiabatic theory. Figure \ref{fig:eigenvaluecrossings} provides an illustration of typical spectral behavior for $H_{N,M}(y)$, suggesting that generally condition \eqref{gapcondition} may only hold locally.

\begin{figure}[ht]
    \centering
    \noindent
    \trimbox{0cm 5.7cm 0cm 0cm}{\begin{tikzpicture}[scale=0.8]
        %axis
        \draw[->] (-1,-0.5)--(10,-0.5);
        \draw[->] (0,-3)--(0,6);
        %ground state eigenvalue E0
        \draw (0.2,6) .. controls (2,-10) and (1,1) .. (10,2);
        %E1
        \draw (0.4,6) .. controls (2,-8) and (5,6) .. (10,3);
        %E2
        \draw (0.6,6) .. controls (3,-8) and (7,6) .. (10,4);
        %E4
        \draw (0.8,6) .. controls (3,-4) and (7,6) .. (5,4);
        %continuous spectrum
        \path(1,6) coordinate(src);
        \path(10,6) coordinate(dst);
        \path(4,2) coordinate(ctrl1);
        \fill[color=white](src) .. controls(ctrl1)  .. (dst)--(10,6)--cycle;
        \draw(src) .. controls(ctrl1)  .. (dst);
        \fill[pattern=dots, very thick](src) .. controls(ctrl1)  .. (dst)--(10,6)--cycle;
        %labels
        \path (10,-0.8) node {$|y|$};
        \path (10,1.6) node {$E_0(y)$};
        \path (10,2.7) node {$E_1(y)$};
        \path (10,3.7) node {$E_2(y)$};
        \path (-1.4,5.7) node {$\sigma(H_{N,M}(y))$};
        \path (5,5.7) node[fill=white] {$\sigma_{\text{ess}}(H_{N,M}(y))$};
    \end{tikzpicture}}
    \caption{An illustration of discrete eigenvalue crossings and eigenvalues merging into the continuous spectrum.}
    \label{fig:eigenvaluecrossings}
\end{figure}

The presence of eigenvalue crossings complicates theories like time-dependent BO and limits their application to chemically relevant cases. A systematic framework characterizing when and where such crossings occur is non-existent. The first rigorous study of molecular Hamiltonian eigenvalue crossings was by Hagedorn \cite{Hag_crossings}, who classified two-level crossings under specific multiplicity conditions (conical intersections). Hagedorn later extended this work to avoided crossings \cite{Hag_avoidedcrossings}, with both studies motivated by understanding molecular propagation through these crossings.

To conclude this introduction we briefly sketch our approach. The total symmetry of the eigenstates and non-degeneracy of the ground state energy are both essential ingredients for the proof. Assume there exists a sequence $\{y^n\}_{n=1}^\infty$ in $\R^{3M}$ such that 
\begin{equation} 
    E_1(y^n) - E_0(y^n) \to 0.
\end{equation}
If $\max_{i \neq j} \nrm{y_i^n - y_j^n}$ is a bounded sequence, then, after translating so that $y_1^n = 0$, $\{y^n\}_{n =1}^\infty$ must be a bounded sequence. Hence there must exist a convergent subsequence $\{y^{n_k}\}_{k=1}^\infty$ such that $y^{n_k} \to y^0$, implying $E(y^0) = E_1(y^0)$ and contradicting the non-degeneracy of the ground state. 
Observe that translations in the $y$-coordinates only affect $H_{N,M}(y)$ via the nuclear-nuclear interaction $V_N(y)$, and shift the spectrum $\sigma(H_{N,M}(y))$ by a constant and do not affect the size of the gap. By translating $y_1^n$ to $0$ and considering $\max_{i \neq j} \{\nrm{y_i^n - y_j^n} \}$ we account for rigid translations of the system (we can also work in center-of-mass coordinates). 

If $\max_{i \neq j} \{\nrm{y_i^n - y_j^n}\}$ is unbounded, we consider a decomposition of the system into non-interacting clusters. The nuclei partition into at least two smaller subgroups which recede from each other, denote this partition of nuclear coordinates by $\mathcal{N}$. Then, the electrons must choose how to partition themselves amongst these subgroups, so we denote the set of all these possible electronic partitions by $\mathcal{A}_{\mathcal{N}}$. 

For a given electronic partition $\mathcal{E} \in \mathcal{A}_{\mathcal{N}}$, we may decompose the operator $H_{N,M}(y^n)$ as
\begin{equation} \label{HNM_HE_IE_cluster_intro}
    H_{N,M}(y^n) = H_{\mathcal{E}}(y^n) + I_{\mathcal{E}}(y^n),
\end{equation}
where $H_{\mathcal{E}}(y^n)$ represents the free Hamiltonian of the decomposed system (clusters with intracluster interactions) and $I_{\mathcal{E}}(y^n)$ the intercluster interactions. Intuitively, we expect $I_{\mathcal{E}}(y^n) = O(1/R_n)$, where $R_n \to \infty$ as $n \to \infty$ is the minimum distance between separate clusters. With \eqref{HNM_HE_IE_cluster_intro} in mind, we expect the eigenvalues of $H_{N,M}(y^n)$ to approach those of $H_{\mathcal{E}}(y^n)$ as $n \to \infty$.

To formalize this, define the new Hilbert space
\begin{equation}
    \mathcal{H}_\infty = \bigotimes_{\mathcal{E} \in \mathcal{A}_\mathcal{N}} L^2(\mathbb{R}^{3N})
\end{equation}
and let
\begin{equation}
    H_\infty = \bigoplus_{\mathcal{E} \in \mathcal{A}_\mathcal{N}} H_{\mathcal{E}},
\end{equation}
where the individual clusters are assumed to reach some sort of fixed configuration in the limit. Let $E_k(H_\infty)$ denote the $k^{th}$ eigenvalue of $H_\infty$ (counting multiplicity), and by $k=\infty$ denote the bottom of the essential spectrum. To list $E_k(H_\infty)$, one can list all of $E_k(H_{\mathcal{E}})$ for each decomposition $\mathcal{E}$ and then combine them into one list, counting multiplicity. Then, we prove that
\begin{equation} \label{E0_E1_limit_introversion}
    \lim_{n \to \infty} E_0(H_{N,M}(y^n)) = E_0(H_\infty), \qquad \lim_{n \to \infty} E_1(H_{N,M}(y^n)) = E_1(H_\infty).
\end{equation}
In particular, consider the system breaking into $k \geqslant 2$ clusters. For the $j^{\text{th}}$ cluster, denote by $E_{j,0}$ the neutral cluster ground state energy, $E_{j,1}$ the neutral cluster first excited state energy, and $E_{j,0}^{\delta_j}$ the ground state energy of the $\delta_j$ ion (i.e. the system having lost or gained $\nrm{\delta_j}$ electrons depending if $\delta_j > 0$ or $\delta_j < 0$, respectively). Then, we also show that
\begin{align}
    E_0(H_\infty) &= \sum_{j=1}^k E_{j,0}(y^0), \label{E0_introversion} \\
    E_1(H_\infty) &= \min_{l, \:\delta_j} \left\{ E_{j,l}(y^0) + \sum_{\substack{j=1 \\ j \neq l}}^k E_{j,0}(y^0), \sum_{j=1}^k E_{j,0}^{\delta_j}(y^0) \right\}. \label{E1_introversion}
\end{align}
Identity \eqref{E0_introversion} follows from Assumption \ref{localneutrality_assumption_introversion}. \eqref{E1_introversion} establishes that the first excited state energy may come from either a configuration of neutral clusters, with all clusters but one being in their ground states and the one cluster in the first excited state, or some configuration of ions (clusters with positive or negative charge) in their ground states. The ion charges $\delta_j$ in the minimum in \eqref{E1_introversion} are subject to natural constraints (such as conservation of the number of electrons). 

In \cite{MS}, Morgan and Simon established \eqref{E0_E1_limit_introversion} for all eigenvalues $E_k(H_\infty)$, $k \geqslant 1$, including $k = \infty$, without assuming \ref{localneutrality_assumption_introversion}. Their analysis, originally for two atoms, generalizes easily to multiple atoms. Here, we explicitly characterize $E_0(H_\infty)$ and $E_1(H_\infty)$ in \eqref{E0_introversion} and \eqref{E1_introversion}, using the time-independent FS map (see Appendix \ref{FSmap_subsection}) and Assumption \ref{localneutrality_assumption_introversion} to define candidate states for eigenstates of the decomposed system.

The FS map, introduced by Feshbach in \cite{Feshbach} for perturbation problems in nuclear reactions, presents an effective Hamiltonian as a Schur complement \cite{Schur1918}. The FS map was applied to eigenvalue perturbation problems in \cite{BFS}, with further refinements in \cite{GD-IMS-BS}. The FS map also featured in the the study of the van der Waals force in \cite{IA-IMS}.

The main advantage of the time-independent FS map is that it transforms the eigenvalue problem of the operator in question, $H_{N,M}(y)$ (which is acting on an infinite dimensional Hilbert space) into an eigenvalue problem for $F_P(\lambda)$ acting on $\Ran P$ (which is a lower dimensional space), where $F_P(\cdot)$ is the FS map, $P$ is an orthogonal projection of finite rank, and $\lambda$ is a spectral parameter. In particular, we choose $P$ to be the orthogonal projection onto the eigenspaces corresponding to $E_0(H_\infty)$ and $E_1(H_\infty)$ (after some consideration and making use of Assumption \ref{localneutrality_assumption_introversion}). Let $P^\perp$ denote the orthogonal projection onto the orthogonal complement. After demonstrating that $P^\perp \left( H_{N,M}(y^n) - E_1(y^n) \right) P^\perp$ is invertible, the FS method states that $\lambda$ is an eigenvalue of $H_{N,M}(y)$ if and only if it is also an eigenvalue of 
\begin{align}
    F_P(\lambda) &= P H_{N,M}(y) P - P H_{N,M}(y) P^\perp (H_{N,M}^\perp(y) - \lambda)^{-1} P^\perp H_{N,M}(y) P, 
\end{align}
acting on the finite dimensional subspace $\textrm{Ran }P$. Hence, the eigenvalue problem
\begin{equation}
    H_{N,M}(y) \psi = \lambda(y) \psi,
\end{equation}
is mapped into the equivalent eigenvalue problem
\begin{equation} \label{FP_lambda_evproblem}
    F_P(\lambda(y)) \phi = \lambda(y) \phi,
\end{equation}
where the latter is lower dimensional and in particular, finite dimensional due to the rank of $P$. In turn, the price paid is that the problem is non-linear (in $\lambda$). Examining the problem \eqref{FP_lambda_evproblem} for $\lambda(y) = E_0(y)$ and $\lambda(y) = E_1(y)$, we obtain expansions for these quantities around $E_0(H_\infty)$ and $E_1(H_\infty)$ respectively, and demonstrate that the errors are $o(1)$ in the limit \eqref{E0_E1_limit_introversion}. Since $E_0(H_\infty)$ must be non-degenerate also, then $E_1(y^n) - E_0(y^n) > 0$ in the limit $n \to \infty$.

\subsection{Outline} Section \ref{ModelandMainResult_Section} is devoted to a detailed introduction and discussion of the model and assumptions. While some of this material is well-known, it is included to establish notation. In particular, in Section \ref{localneutrality_subsection} we precisely re-state Assumption \ref{localneutrality_assumption_introversion}, and in Section \ref{mainresult_proof_subsection} we precisely re-state the main theorem and give the proof. The proof is dependent on the eigenvalue limits which we establish in Section \ref{eigenvaluelimits_section}. In Appendix \ref{FSmap_subsection} we recall the time-independent FS map theory and give further references.

%%%%%%%%%%%%%%%%%%%%%%%%%%%%%%%%%%%%%%%%%%%%%%%%%%%%%%%%%%%%%%%%%%%%%%%%%%%
\section{Model and Main Result} \label{ModelandMainResult_Section}
%%%%%%%%%%%%%%%%%%%%%%%%%%%%%%%%%%%%%%%%%%%%%%%%%%%%%%%%%%%%%%%%%%%%%%%%%%%

\subsection{The Schr{\"o}dinger Hamiltonian for a molecule} \label{SE_operator_subsection}
We consider a neutral molecule as described by the Hamiltonian $H_{N,M}(y)$, given in \eqref{Hbo(y)_withCoulomb} - \eqref{neutralmolecule}. We consider the operator as acting on the full state space $L^2(\mathbb{R}^{3N})$ and introduce the symmetry subspace later, in Subsection \ref{particlestatistics_subsection}. The results that follow are independent of particle statistics. We ignore spin.

Recall we denote by $E_0(y)$ the bottom of the spectrum of $H_{N,M}(y)$ (cf. \eqref{E0_def_intro}). When $N < Z + 1$ (corresponding to neutral or positively charged molecules), where $Z = \sum_{j=1}^M Z_j$ is the total nuclear charge, the celebrated Hunziker-van Winter-Zhislin (HVZ) and Zhislin-Sigalov theorems state that $E_0(y) \in \sigma_d(H_{N,M}(y))$, lying strictly below the essential spectrum. 

\begin{theorem} \label{Hbospectrumcontinuous}
(a) (Hunziker-van Winter-Zhislin) We have 
\begin{equation}
    \sigma_{ess}(H_{N,M}(y)) = [\Sigma(y), \infty)
\end{equation}
where $\Sigma(y) := \inf \sigma(H_{N-1,M}(y))$ and $H_{N-1,M}(y)$ is the Born-Oppenheimer Hamiltonian of a molecule with one less electron. The value $\Sigma(y)$ is also known as the ionization threshold for $H_{N,M}(y)$. \\
(b) (Zhislin-Sigalov) Let $N < \sum_{j=1}^M Z_j + 1$ (i.e. neutral or positively charged molecules). Then, the operator $H_{N,M}(y)$ has an infinite number of eigenvalues $E_j$, for $j = 0, 1, \dots$ counting multiplicity, below its ionization threshold $\Sigma(y)$.
\end{theorem}

By a bound state we mean a $L^2$-integrable eigenfunction of $H_{N,M}$ corresponding to an energy below the continuum. Zhislin and Sigalov showed that any neutral atom or positive ion has infinitely many bound states, and this was extended to molecules as well. Negative ions may have at most finitely many bound states below the continuum (see \cite{Vugalter1992, IMS-1}), and the maximum negative ionization of a molecule is bounded. For our purposes it will be sufficient to refer to the result of Lieb \cite{EL-1}, as follows\footnote{Although this was improved by Nam in \cite{Nam2012} for $Z \geqslant 6$, we do not require this improvement.}. The maximum number of electrons $N_c$ that a molecule with $M$ nuclei and total nuclear charge $Z$ can bind (i.e. $H_{N_c, M}$ has a well-defined normalizable ground state) satisfies the inequality
\begin{equation} \label{Lieb_nonbinding}
    N_c < 2 Z + M.
\end{equation}
This, for example, implies the instability of the di-anion \ce{H^2-}.

\begin{remark}
The theorized optimal upper bound for the number of electrons is $N_c < Z + C M$, for some constant $C > 0$ (possibly $C = 1$) and is known as the \textit{ionization conjecture}. This is still an open problem (see \cite{PTN-2} for an overview and discussion), and under fermionic statistics there is much numerical \cite{Hogreve1998, SergeevSabre-1} and experimental \cite{AHH-1} evidence. In particular, it has been shown to depend crucially on the Pauli exclusion principle\footnote{In the bosonic case it was shown by Lieb and Benguria \cite{PhysRevLett.50.1771},
\begin{equation}
    \liminf_{Z \to \infty} \frac{N_c(Z)}{Z} \geqslant t_c > 1,
\end{equation}
and the constant $t_c \approx 1.21$ is known (\cite{Baumgartner_1984}).}. While in \cite{Hogreve2011} it was proven directly that the bosonic anion \ce{He-} can exist as a stable atom, in \cite{Goto2018} the nonexistence of bosonic anions was shown under the assumption of zero binding energy.
\end{remark}

It is possible to have eigenvalues embedded in the continuum - the theory of resonances - but these are generally expected to be unstable under perturbations (see for example \cite{MM-IMS}). Physically, continuum states should be states which are unstable with respect to break-up, capable of separating into at least $2$ spatially separated clusters. The HVZ Theorem suggests that the essential (continuous) spectrum of $H_{N,M}(y)$ originates from the molecule shedding an electron which moves freely at infinity. 

The next theorem, by Combes and Thomas, states that $H_{N,M}(y)$ has spatially well-localized eigenstates. 

\begin{theorem} \label{Hbospectrumdiscrete}
(Combes-Thomas) The eigenfunctions $\psi_j(y)$ corresponding to the eigenvalues $E_j(y)$ of $H_{N,M}(y)$ below the continuum are exponentially localized, i.e. for $E_j(y) < \Sigma(y)$ there exists a constant $\sqrt{\Sigma(y) - E_j(y)} > \alpha_j > 0$ such that
\begin{equation} \label{exponentiallocalizationofeigenfunctions}
    \lVert{e^{\alpha_j \langle{x-z}\rangle} \partial^\theta_x \tau_z \psi_j(y)}\rVert < \infty,
\end{equation}
for any multi-index $0 \leqslant \lvert{\theta}\rvert \leqslant 2$. Here $\langle{x}\rangle = (1 + \lvert{x}\rvert^2)^{1/2}$ and $\tau_h$ denotes the translation by $h \in \R^{3N}$ in the $x$-coordinate, i.e. $\tau_h \psi_j(y)(x) = \psi_j(y)(x-h)$.
\end{theorem}

Both Theorem \ref{Hbospectrumcontinuous} and Theorem \ref{Hbospectrumdiscrete} are well-known and can be found in many references, such as \cite{GS, HS, AL, IA-IMS, BS-1} and references therein.

We remark that $E_0(y)$ is a translation-invariant function of the nuclear positions $y$, and $y \mapsto E_0(y)$ is known to be continuous away from nuclear overlap (and even analytic, see \cite{Combes-Seiler-1978} for a diatomic case and \cite{RSv4} for a perturbation theory argument). For the claim of continuity we quickly sketch an elementary argument. Let $\psi_0(y)$ denote the ground state, i.e. $H_{N,M}(y) \psi_0(y) = E_0(y) \psi_0(y)$. For $h \in \R^{3M}$ with $\nrm{h}$ small, 
\begin{align}
    E_0(y) = \langle \psi_0(y), H_{N,M}(y) \psi_0(y) \rangle \leqslant \langle \psi_0(y+h), H_{N,M}(y) \psi_0(y+h) \rangle.
\end{align}
Similarly for $E_0(y+h)$, 
\begin{align}
    E_0(y+h) &= \langle \psi_0(y+h), H_{N,M}(y+h) \psi_0(y+h) \rangle \leqslant \langle \psi_0(y), H_{N,M}(y+h) \psi_0(y) \rangle.
\end{align}
As $H_{N,M}(y) - H_{N,M}(y+h)$ is a difference of Coloumb potentials, which are relatively bounded by the Laplacian with relative bound zero, we write $H_{N,M}(y) - H_{N,M}(y+h) = o_{H^2 \to L^2}(1)$ in the limit as $\nrm{h} \to 0$, and thus
\begin{align}
    \langle \psi_0(y+h), H_{N,M}(y) \psi_0(y+h) \rangle &\approx E_0(y+h) + o(1), \\
    \langle \psi_0(y), H_{N,M}(y+h) \psi_0(y) \rangle &\approx E_0(y) + o(1).
\end{align}
Combining the above two equations, we have $E_0(y) - E_0(y+h) = o(1)$ in the limit as $\nrm{h} \to 0$, so the claim follows.

Let $R$ be the minimum distance between any two nuclei, 
\begin{equation}
    R(y) = \min \{\lvert{y_i - y_j}\rvert : i,j \in \{1, \dots, M\}, i \neq j\}. \label{leastdistbetweennuclei}
\end{equation}
The following Lemma \ref{Rto0limitlemma} is elementary but important. The first part establishes that $E_0 \to \infty$ when $R \to 0$, and so we adopt the convention that $E_0(y) = +\infty$ when two nuclei are on top of each other. In the second part, we establish that despite this blowup, the spectral gap is preserved, in the following sense. Let $\tilde{H}_{N,M}(y) = H_{N,M}(y) - V_n(y)$, i.e. the electronic Hamiltonian without the nuclei-nuclei interaction. Denote by $\tilde{E}_j(y)$ for $j=0, 1, \dots$ the corresponding eigenvalues of $\tilde{H}_{N,M}(y)$, counting multiplicities. Heuristically $V_n = O(1/R)$, but for fixed $y$, $V_n$ only serves to raise the eigenvalues by a constant, so the spectral gap is independent of $V_n$.

\begin{lemma} \label{Rto0limitlemma}
a) In the limit as $R \to 0$, 
\begin{equation}
    E_j(y) \to \infty. \label{E^j(y)toinfty}
\end{equation}
b) We have that
\begin{equation} \label{gapconditionindependentofR}
    E_1(y) - E_0(y) = \tilde{E}_1(y) - \tilde{E}_0(y).
\end{equation}
\end{lemma}

This lemma is standard and we omit the proof. 

% For the convenience of the reader we provide a proof.

% \begin{proof}[Proof of Lemma \ref{Rto0limitlemma}]
% a) It is sufficient to prove that $E_0(y) \to \infty$, as it lies below the rest of the spectrum. By form boundedness of the Coulomb potentials, there exists some finite constant $C$, independent of $y$, such that 
% \begin{equation}
%     H_{N,M}(y) \geqslant V_n(y) - C. \label{boundonE^1}
% \end{equation}
% This implies $E_0(y) \geqslant V_n(y) - C$ and since $V_n(y) \to \infty$ as $R \to 0$ we obtain (\ref{E^j(y)toinfty}).

% b) Since $V_n = V_n(y)$ is independent of $x$, it acts as a constant in the eigenvalue problem for $H_{N,M}$. We have 
% \begin{align}
%      E_0(y) &= \min_{\lVert{\psi}\rVert_{L^2_x} = 1} \langle \psi, H_{N,M}(y) \psi \rangle_{L^2_x} = \tilde{E}_0(y) + V_n(y)
% \end{align}
% and also 
% \begin{equation}
%     \tilde{H}_{N,M}(y) \psi_\circ(y) = (E_0(y) - V_n(y))\psi_\circ(y),
% \end{equation}
% so $\tilde{\psi}_\circ = \psi_\circ$. Then analogously we have $E_1(y) = \tilde{E}_1(y) + V_n(y)$. From this (\ref{gapconditionindependentofR}) follows. 
% \end{proof}

\begin{remark}
If we regularize the Coulomb interaction potentials by replacing the point nuclear charges with smeared ones, then $V_{en}$ and $V_n$ are bounded and $x \to V_{en}(x, \cdot) \in L^\infty_y$ is smooth. In this case Lemma \ref{Rto0limitlemma} (b) holds but Lemma \ref{Rto0limitlemma} (a) does not. Instead, using perturbation techniques it can be shown that
\begin{equation}
    \lVert{E_j}\rVert_{L^\infty_y} < \infty.
\end{equation}
\end{remark}

\subsection{Cluster decompositions} In general, a breakup of a molecule into smaller molecules or atoms is denoted by a cluster decomposition
\begin{equation}
    \mathcal{N}\mathcal{E} := \big\{ \{\mathcal{N}_1,\mathcal{E}_1\}, \dots, \{\mathcal{N}_k, \mathcal{E}_k\} \big\},
\end{equation}
where $\mathcal{N} = \{\mathcal{N}_1, \dots, \mathcal{N}_k\}$ is a partition of the nuclear coordinates $\{1, \dots, M\}$ into disjoint subsets such that $\sqcup_{j=1}^k \mathcal{N}_j = \{1, \dots, M\}$ and $\mathcal{E} = \{\mathcal{E}_1, \dots, \mathcal{E}_k\}$ is a partition of the electron coordinates $\{1, \dots, N\}$ into disjoint subsets such that $\sqcup_{j=1}^k \mathcal{E}_j = \{1, \dots, N\}$. 

Let $\mathcal{A}$ denote the set of all possible cluster decompositions $\mathcal{N}\mathcal{E}$. Intuitively, as the system breaks into clusters, electrons must choose a nuclei to stay with. Then, let
\begin{equation}
    \mathcal{A}_\mathcal{N} = \big\{ \mathcal{E} = \{\mathcal{E}_1, \dots, \mathcal{E}_k\} \:|\: \textrm{ such that } \mathcal{N}\mathcal{E} \in \mathcal{A} \big\},
\end{equation}
i.e. $\mathcal{A}_\mathcal{N}$ is the set of all possible distributions of electrons amongst the clusters defined by the groups of nuclei according to the fixed partition $\mathcal{N}$. 

Consider an unbounded sequence $\{y^n\}_{n=1}^\infty$ such that $\max_{i \neq j} \{\nrm{y_i^n - y_j^n} \} \to \infty$. This corresponds to the many body system breaking into at least two smaller clusters which are each localized around their center of mass. Let $B(z,C)$ denote the ball of radius $C$ centered at $z$. For the sequence $\{y^n\}_{n=1}^\infty$ we define an \textit{admissible partition of the nuclei into $k \geqslant 2$ separate clusters} to be a partition $\mathcal{N}$ together with a sub-sequence of $\{y^n\}_{n=1}^\infty$ which we again denote by $\{y^n\}_{n=1}^\infty$ such that for all nuclei $i \in \mathcal{N}_j$, 
\begin{equation}
    y_i^n = z_j^n + Y_i^n, \qquad \text{ for all } j,
\end{equation}
and where for the cluster $\mathcal{N}_j$ there exists a constant $C_j > 0$ such that $y_i^n \in B(z_j^n, \frac{C_j}{2})$, and therefore $\lvert{Y_l^n - Y_k^n}\rvert \leqslant C_j$ for all $l, k$ in the cluster. We refer to $Y_i^n$ as the cluster coordinates of the nucleus $i$, and to $z_j^n$ as the center of the cluster $\mathcal{N}_j$. The sequences $\{Y_i^n\}_n$ are bounded for all $i$, while $\{z_j^n - z_i^n\}_n$ are unbounded for all $i \neq j$, and we choose a convergent subsequence along which we write $Y_i^n \to Y_i^0$. We refer to this as \textit{each cluster reaching a fixed internal geometry}. For $i \neq j$ belonging to different clusters, $\lvert{y^n_i - y^n_j}\rvert \to \infty$.

Denote by $y_{\mathcal{N}_j}^n$ the tuple of all $y$-coordinates with indices belonging to $\mathcal{N}_j$, and similarly for the cluster coordinates $Y_{\mathcal{N}_j}^n$. For $\mathcal{E} \in \mathcal{A}_\mathcal{N}$, we also denote by $x_{\mathcal{E}_j}$ the tuple of all $x$-coordinates with indices belonging to $\mathcal{E}_j$. 

\begin{lemma} \label{admissiblepartitionlemma}
Consider a system described by $H_{N,M}(y)$ given by \eqref{Hbo(y)_withCoulomb} - \eqref{neutralmolecule}, from which we drop the potential $V_n$. Then, for a sequence $\{y^n\}_{n=1}^\infty$ with $\max_{i \neq j} \{\nrm{y_i^n - y_j^n} \} \to \infty$ for at least one pair $i \neq j$, an admissible partition $\mathcal{N}$ of the nuclei exists and each cluster reaches a fixed internal geometry. There must be at least $k=2$ clusters, each containing at most $M-1$ nuclei.

Furthermore, in the corresponding cluster system of coordinates, $H_{N,M}(y^n)$ takes the form
\begin{equation} \label{Hbo=Hcluster+Interaction}
    H_{N,M}(y^n) = H_{\mathcal{E}}(Y^n) + I_{\mathcal{E}}(Y^n),
\end{equation}
where
\begin{equation}
    H_{\mathcal{E}} = \sum_{j=1}^k H_{\mathcal{E}_j}(Y_{\mathcal{N}_j}^n), \quad I_{\mathcal{E}} = \sum_{1 \leqslant i < j \leqslant k} I_{\mathcal{E}_i, \mathcal{E}_j}(Y_{\mathcal{N}_i}^n, Y_{\mathcal{N}_j}^n). \label{Hboclusters}
\end{equation}

In \eqref{Hboclusters}, $H_{\mathcal{E}_j}$ denotes the Born-Oppenheimer Hamiltonian of the system of static nuclei of indices in $\mathcal{N}_j$ at fixed positions $y_{\mathcal{N}_j}^n$ and the electrons with indices in $\mathcal{E}_j$ interacting between themselves and with each other. By $I_{\mathcal{E}_i, \mathcal{E}_j}$ we denote all of the interaction terms between the different clusters $i \neq j$.

Define $r_{ij}^n = z_i^n - z_j^n$ to be the separation between the clusters $\mathcal{N}_i$ and $\mathcal{N}_j$. Clearly $\lvert{r_{ij}^n}\rvert \to \infty$. Then, shifting the coordinates of the electrons to their corresponding cluster, $\tilde{x}_l = x_l - z_j^n$, where $x_l \in \mathbb{R}^3$ (and we drop the dependence on $n$ in the coordinate), we have
\begin{align}
    &H_{\mathcal{E}_j}(Y_{\mathcal{N}_j}^n) = \tau_n^{-1} \left(\sum_{l \in \mathcal{E}_j} -\frac{1}{2m_e} \Delta_{\tilde{x}_l^n} + \frac{1}{2} \sum_{\substack{l,r \in \mathcal{E}_j \\ l \neq r}} \frac{e^2}{\lvert{\tilde{x}_l - \tilde{x}_r}\rvert} - \sum_{l \in \mathcal{E}_j} \sum_{r \in \mathcal{N}_j} \frac{e^2 Z_r}{\lvert{\tilde{x}_l - Y_r^n}\rvert} \right) \tau_n, \label{Hcluster}\\
    &I_{\mathcal{E}_i, \mathcal{E}_j}(Y_{\mathcal{N}_i}^n, Y_{\mathcal{N}_j}^n) = \tau_n^{-1} \left(\frac{1}{2} \sum_{l \in \mathcal{E}_i} \sum_{s \in \mathcal{E}_j} \frac{e^2}{\lvert{\tilde{x}_l - \tilde{x}_s + r_{ij}^n}\rvert} - \sum_{l \in \mathcal{E}_i} \sum_{s \in \mathcal{N}_j} \frac{e^2 Z_s}{\lvert{\tilde{x}_l - Y_s^n + r_{ij}^n}\rvert}\right. \nonumber \\
    &\hspace{88pt}\left. - \sum_{l \in \mathcal{N}_i} \sum_{s \in \mathcal{E}_j} \frac{e^2 Z_l}{\lvert{Y^n_l - \tilde{x}_s + r_{ij}^n}\rvert} \right) \tau_n. \label{clusterinteraction}
\end{align}
Here, we define the translation $\tau_n$ to be 
\begin{equation} \label{clustertranslation}
    \tau_n \phi(x_1, \dots, x_N) = \phi(x_1 - \delta^n_1, \dots, x_N - \delta^n_N), \qquad \delta^n_i = z_j^n \:\text{ for }\: i \in \mathcal{E}_j.
\end{equation}
The operators defined in \eqref{Hcluster} and \eqref{clusterinteraction} above act in the shifted coordinates $\tilde x$, and in particular we have $\phi(\tilde x) = \tau_n \phi(x)$. We adopt the convention that operators parameterized in the $y^n$ coordinates act on untranslated coordinates (functions $\phi(x)$) while operators parameterized in the $Y^n = (Y^n_{\mathcal{N}_1}, \dots, Y^n_{\mathcal{N}_k})$ coordinates act on translated coordinates (functions $\phi(\tilde x)$). 

A system containing ions is characterized by 
\begin{equation} \label{ioncharges}
    I = \left\{\nrm{\mathcal{E}_j} - \tilde{Z}_j  \right\}_{j=1}^k, \qquad \tilde{Z}_j = \sum_{i \in \mathcal{N}_j} Z_i,
\end{equation}
where $\tilde{Z}_j$ is the total nuclear charge of the $j^{\text{th}}$ cluster. We will only invoke this notation in the case where $I$ is not identically $0$ (i.e. at least two ions exist). If $\delta_j \equiv \nrm{\mathcal{E}_j} - \sum_{i \in \mathcal{N}_j} Z_i > 0$, then the cluster $j$ has lost $\delta_j$ electrons as is called a positive ion, cation, or $+\delta_j$ ion. If $\delta_j < 0$, then the system has gained $\nrm{\delta_j}$ electrons and is called a negative ion, anion, or $-\nrm{\delta_j}$ ion. 
\end{lemma} 

\begin{figure}[ht]
    \centering
    \begin{tikzpicture}[scale=1]
        \draw[dotted] (0,0) circle (40pt);
        \draw[dotted] (-4,3) circle (50pt);
        \draw[dotted] (2,4) circle (60pt);

        \node at (1.6,-1) [xshift=0.1cm, yshift=-0.25cm] {$B(0, \frac{C_1}{2})$};
        \node at (-5,1.2) [xshift=0cm, yshift=-0.25cm] {$B(z^n_2, \frac{C_2}{2})$};
        \node at (3.5,2) [xshift=0cm, yshift=-0.25cm] {$B(z^n_3, \frac{C_3}{2})$};

        \filldraw (0,0) circle (1pt);
        \node (zn1) at (0,0) {};
        \node at (0,0) [xshift=0.1cm, yshift=-0.25cm] {$z^n_1$};
        \filldraw (-4,3) circle (1pt);
        \node (zn2) at (-4,3) {};
        \node at (-4,3) [xshift=0.1cm, yshift=-0.25cm] {$z^n_2$};
        \filldraw (2,4) circle (1pt);
        \node (zn3) at (2,4) {};
        \node at (2,4) [xshift=0.1cm, yshift=-0.25cm] {$z^n_3$};

        \draw[->, red] (zn2) -- (-4.5,3.5);
        \draw[->, red] (zn3) -- (2.5,4.5);

        \filldraw (-0.5,-0.8) circle (1pt);
        \node (Yn1) at (-0.5,-0.8) {};
        \node at (-0.5,-0.8) [xshift=0.1cm, yshift=-0.3cm] {$Y^n_1$};
        \filldraw (-0.2,1) circle (1pt);
        \node (Yn2) at (-0.2,1) {};
        \node at (-0.2,1) [xshift=0.35cm, yshift=-0.25cm] {$Y^n_2$};
        \draw[dotted] (Yn1) -- (zn1);
        \draw[dotted] (Yn2) -- (zn1);

        \filldraw (-3.5,2) circle (1pt);
        \node (Yn3) at (-3.5,2) {};
        \node at (-3.5,2) [xshift=-0.15cm, yshift=-0.3cm] {$Y^n_3$};
        \filldraw (-3.8,4) circle (1pt);
        \node (Yn4) at (-3.8,4) {};
        \node at (-3.8,4) [xshift=-0.2cm, yshift=0.25cm] {$Y^n_4$};
        \draw[dotted] (Yn3) -- (zn2);
        \draw[dotted] (Yn4) -- (zn2);

        \filldraw (1,5) circle (1pt);
        \node (Yn5) at (1,5) {};
        \node at (1,5) [xshift=0.2cm, yshift=0.25cm] {$Y^n_5$};
        \filldraw (1.5,3) circle (1pt);
        \node (Yn6) at (1.5,3) {};
        \node at (1.5,3) [xshift=0.3cm, yshift=-0.25cm] {$Y^n_6$};
        \draw[dotted] (Yn5) -- (zn3);
        \draw[dotted] (Yn6) -- (zn3);

        \draw[<->, blue] (Yn3) -- (Yn2) node [midway, below=2.5pt, fill=white] {$r^n_{12}$};
        \draw[<->, blue] (Yn2) -- (Yn6) node [midway, right=5.5pt] {$r^n_{13}$};
        \draw[<->, blue] (Yn4) -- (Yn5) node [midway, above=2.5pt, fill=white] {$r^n_{23}$};
    \end{tikzpicture}
    \caption{An example of the cluster decomposition of a system of $6$ nuclei with labelled cluster coordinates $Y_j^n$, $j = 1, \dots, 6$, into $3$ clusters of centers $z^n_1, z^n_2, z^n_3$. The cluster distances are labelled $r^n_{12}, r^n_{13}$, and $r^n_{23}$.}
    \label{fig:clusterdecomp}
\end{figure}

\begin{proof}[Proof of Lemma \ref{admissiblepartitionlemma}]
Once we have obtained an admissible partition and cluster system of coordinates, the expressions \eqref{Hbo=Hcluster+Interaction} - \eqref{clusterinteraction} follow immediately by making the suitable change of coordinates.

Assume that $y^n_1 = 0$, else without loss of generality we translate the system (which does not change the spectrum). We also choose our labelling such that the index $1$ belongs to $\mathcal{N}_1$. In the case of having only two nuclei, the only admissible partition is $\mathcal{N}_1 = \{1\}$ and $\mathcal{N}_2 = \{2\}$, where $\lvert{y^n_2}\rvert \to \infty$. For three nuclei, 
\begin{enumerate}
    \item If $y^n_2$ has a bounded sub-sequence $y^{n_k}_2$, we put nucleus $2$ in the same cluster as nucleus $1$ and work with the sub-sequence. Denote the bounded subsequence $y^{n_k}_2$ by $y^n_2$. 
    \begin{enumerate} 
        \item If the subsequence $y^n_3$ has a bounded subsubsequence $y^{n_k}_3$, we put nucleus $3$ in the same cluster as nuclei $1$ and $2$ and work with the subsequence. Denote the bounded subsequence $y^{n_k}_3$ by $y^n_3$. In this case we have $\mathcal{N}_1 = \{1, 2, 3\}$, which is unattainable in the three nuclei case, but becomes relevant in the cases of four nuclei or more. 
        \item $\lvert{y^n_3}\rvert \to \infty$ so the nuclei $1$ and $2$ are in the same cluster, while the nucleus $3$ is in a different cluster. Then we have $\mathcal{N}_1 = \{1, 2\}$ and $\mathcal{N}_2 = \{3\}$.
    \end{enumerate}
    \item Else, $\lvert{y^n_2}\rvert \to \infty$ so the nuclei $1$ and $2$ are in different clusters. 
    \begin{enumerate} 
        \item If $y^n_3$ has a bounded subsequence $y^{n_k}_3$, we put nucleus $3$ in the same cluster as nucleus $1$, while nucleus $2$ is in a different cluster. We again work with the subsequence and denote the bounded subsequence $y^{n_k}_3$ by $y^n_3$. Then $\mathcal{N}_1 = \{1, 3\}$ and $\mathcal{N}_2 = \{2\}$. 
        \item If $\lvert{y^{n}_2 - y^{n}_3}\rvert$ has a bounded subsequence, we put nucleus $3$ in the same cluster as nucleus $2$ and nucleus $1$ is in its own cluster. Then $\mathcal{N}_1 = \{1\}$ and $\mathcal{N}_2 = \{2, 3\}$. 
        \item None of the above occur, i.e. $\lvert{y^n_3}\rvert \to \infty$ and $\lvert{y^n_2 - y^n_3}\rvert \to \infty$, then all three nuclei $1, 2$ and $3$ are in separate clusters. So we have $\mathcal{N}_1 = \{1\}$, $\mathcal{N}_2 = \{2\}$, and $\mathcal{N}_3 = \{3\}$.  
    \end{enumerate}
\end{enumerate}

We iterate this comparison each time we add a nuclei. Given that $k-1$ nuclei have been partitioned in an admissible way, we compare the sequence $y^n_k$ of the $k^{th}$ nucleus with every coordinate $y^n_1, \dots, y^n_{k-1}$. If for some coordinate $j \in \{1, \dots, k-1\}$ we have a sub-sequence $n_l$ such that $\lvert{y^{n_l}_j - y^{n_l}_k}\rvert$ is uniformly bounded, then the nucleus $k$ is placed into the same cluster as nucleus $j$. Else, it becomes the first element of a new cluster. In this manner the clusters are all well-defined. 

Thus, we have constructed a subsequence of $y^n$ which we also denote by $y^n$ and a partition $\mathcal{N} = \{\mathcal{N}_1, \dots, \mathcal{N}_k\}$ of $\{1, \dots, M\}$, such that $\lvert{y^n_i - y^n_j}\rvert$ is uniformly bounded for all $i, j$ in the same cluster, and $\lvert{y^n_i - y^n_j}\rvert \to \infty$ otherwise. 

It remains to modify the subsequence such that each cluster attains a fixed internal geometry. Without loss of generality, consider the cluster $\mathcal{N}_1$ containing $y_1, \dots, y_l$. Since $y^n_1 - y^n_2$ is bounded, we work with the subsequence indexed by $n_k$, such that $y^{n_k}_1 - y^{n_k}_2$ converges to a fixed vector. On this subsequence we take a subsubsequence $n_{k_r}$ such that $y^{n_{k_r}}_1 - y^{n_{k_r}}_3$ converges to a fixed vector. Iterate this process for all the nuclei in $\mathcal{N}_1$, and then for all nuclei in $\mathcal{N}_2$, and so on; and we index the corresponding sub...subsequence by $n$.

We thus obtain a sequence indexed by $n$ such that (for sufficiently large $n$) there exists a constant $\tilde{C}_j > 0$ for the cluster $\mathcal{N}_j$ such that for all nuclear indices $i,k$ belonging to the cluster, $\lvert{y_i^n - y_k^n}\rvert \leqslant \tilde{C}_j$. Then, there exists a point $z_j^n$ such that $y_i^n \in B(z_j^n, C_j)$ for all $i$ in the cluster $\mathcal{N}_j$ (the center of mass of the cluster). In this case we define $Y_i^n = y_i^n - z_j^n$, and choosing a subsequence such that $Y_i^n$ converges to $Y^0_i$, we observe that $Y_i^n$ and $z_j^n$ have the desired properties.
\end{proof}

\subsection{Local neutrality} \label{localneutrality_subsection}
As described in Section \ref{Introduction_Section}, in \cite{MS} Morgan and Simon proved that in the decomposition limit the energy of the molecule becomes the sum of the energies of the sub-systems. For the ground state we may write
\begin{equation} \label{eigenvalue_clusterlimit_MS}
    \lim_{n \to \infty} E_0(H_{N,M}(y^n)) = \min_{\mathcal{E} \in \mathcal{A}_\mathcal{N}} \left\{ \sum_{j=1}^k E_0(H_{\mathcal{E}_j}(Y^0_{\mathcal{N}_j})) \right\},
\end{equation}
where recall $Y^0_{\mathcal{N}_j}$ is the limit of the $y$-coordinates in cluster center of mass coordinates.

It is a famous conjecture that the minimum on the right-hand side of \eqref{eigenvalue_clusterlimit_MS} is attained only for the neutral configurations, i.e. the $\mathcal{E} \in \mathcal{A}_\mathcal{N}$ for which $\lvert\mathcal{E}_j\lvert = \sum_{i \in \mathcal{N}_j} Z_i$ (in everyday experience, most objects encountered are neutral). We assume this to be true in the Assumption \ref{localneutrality_assumption}.

\begin{assumption}{[LNc]} \label{localneutrality_assumption}
(Local neutrality for clusters.) Given a cluster decomposition of the nuclei, the electrons divide themselves amongst the clusters so that each cluster is neutral,
\begin{equation}
    \nrm{\mathcal{E}_j} = \tilde{Z}_j \quad \forall j = 1, \dots, k,
\end{equation}
i.e. given a partition of the nuclei $\mathcal{N}$,
\begin{equation} \label{localneutrality_minimizer}
    \min_{\mathcal{E} \in \mathcal{A}_\mathcal{N}} \left\{ \sum_{j=1}^k E_0(H_{\mathcal{E}_j}) \right\} = \sum_{j=1}^k E_0(H_{\overline{\mathcal{E}}_j}),
\end{equation}
where the minimizing partition $\overline{\mathcal{E}}$ is only attained by neutral clusters. (The minimizer is not unique, as any minimizer remains a minimizer under any permutation of electrons that preserves the clusters.)
\end{assumption}

It was established by in \cite{LT} that all neutral molecules can bind in the BO approximation, meaning that there exists a minimum of the ground state energy $E_0(y)$ with respect to the nuclear positions. 

It was shown in \cite{IA-IMS, anapolitanos2016} that Assumption \ref{localneutrality_assumption} holds for a system of hydrogen atoms, see the discussions surrounding property (E) contained in \cite[Appendix A]{IA-IMS} and \cite[Proposition 3.1]{anapolitanos2016}. For other systems of atoms (and molecules), it is an open problem if local neutrality holds.

% \begin{theorem} \label{neutralmoleculescanbind_theorem}
% (Theorem \ref{neutralmoleculescanbind_theorem} in \cite{AL}) For a neutral molecule ($N = \sum_{j=1}^M Z_j$) there exist some nuclear positions $\overline{y} \in \mathbb{R}^{3M}$ such that
% \begin{equation}
%     E_0(\overline{y}) = \inf_{y \in \mathbb{R}^{3M}} E_0(y).
% \end{equation}
% This follows from the estimate
% \begin{equation}
%     \min_{y \in \mathbb{R}^{3M}} E_0(y) < \liminf_{R \to \infty} E_0(y)
% \end{equation}
% where recall $R$ was defined in (\ref{leastdistbetweennuclei}) to be the shortest distance between nuclei. 
% \end{theorem}

\subsection{Particle statistics}\label{particlestatistics_subsection} None of the results mentioned in the two previous subsections depend on the statistics of particles, or on spin. However, as we remark at the end of this section, the particle statistics play a very important role in the spectral gap result. 

Let $S_N$ denote the set of permutations of the coordinates $\{1, \dots, N\}$ and let $\pi$ be an element of $S_N$. Let $T_\pi$ be a unitary representation of $S_N$ on $L^2(\mathbb{R}^{3N})$, defined through
\begin{equation}
    (T_\pi \psi)(x_1, \dots, x_N) = \psi(x_{\pi^{-1}(1)}, \dots, x_{\pi^{-1}(N)}).
\end{equation}
For $\pi \in S$ we denote by $\lvert{\pi}\rvert$ the number of transpositions making up the permutation $\pi$, so that $(-1)^{\lvert{\pi}\rvert}$ is the parity of $\pi \in S_n$. Then we denote the subspaces of $L^2(\mathbb{R}^{3N})$ for bosons (totally symmetric wavefunctions) and fermions (totally anti-symmetric wavefunctions) respectively as
\begin{align}
    \mathcal{H}_{s} &= \{\psi \in L^2(\mathbb{R}^{3N}) \::\: T_\pi \psi = \psi \:\:\:\forall \pi \in S_N\}, \label{Hilbert_symmetric} \\ 
    \mathcal{H}_{a} &= \{\psi \in L^2(\mathbb{R}^{3N}) \::\: T_\pi \psi = (-1)^{\lvert{\pi}\rvert}\psi \:\:\:\forall \pi \in S_N\}. \label{Hilbert_antisymmetric}
\end{align}
Futhermore, let $S$ denote the symmetrizer and $A$ the anti-symmetrizer respectively, 
\begin{equation} \label{def_symmetrizer}
    S = \frac{1}{N!} \sum_{\pi \in S_N} T_\pi, \qquad A = \frac{1}{N!} \sum_{\pi \in S_N} (-1)^{\lvert{\pi}\rvert} T_\pi.
\end{equation}
Recall that $S$ and $A$ are orthogonal projections on $L^2(\R^{3N})$, i.e. $S^2 = S$, $A^2 = A$, and $S A = 0 = A S$. 

We will denote by $H^S_{N,M}$ or $H^A_{N,M}$ the operator $H_{N,M}$ acting on the corresponding symmetry subspace $\cH_s$ or $\cH_a$ respectively. More specifically, 
\begin{equation}
    H^S_{N,M}(y) = S H_{N,M}(y) S |_{\textrm{Ran }S}, \qquad H^A_{N,M}(y) = A H_{N,M}(y) A |_{\textrm{Ran }A}.
\end{equation}
In general $\sigma(H^S_{N,M}(y)) \neq \sigma(H_{N,M}(y))$ as operators acting on $L^2(\mathbb{R}^{3N})$, but the two operators have the same ground state as the ground state energy of $H_{N,M}(y)$ is non-degenerate, positive, and $[S, H_{N,M}] = 0$. (Hence if $H_{N,M} \psi = \lambda \psi$ then $H_{N,M} S \psi = \lambda S \psi$. As $\psi > 0$, then $S \psi > 0$. Therefore $S \psi$ must also be an eigenfunction and uniqueness holds, see \cite{RSv4}.) Moreover, the symmetry ensures that if the system decomposes into neutral non-interacting clusters, the ground state of the decomposed system is still non-degenerate. This is a key point in the proof of the main result and we illustrate it with the simple case of the gas $H_2$. 

\begin{example}
Consider two hydrogen atoms consisting of two hydrogen nuclei at positions $y_1$ and $y_2$ and let $x_1$ and $x_2$ denote the position coordinates of the electrons. We consider the simplest possible limit by changing coordinates, so that $y_1 = 0$ and fixing a unit vector $\hat{e} \in \mathbb{R}^3$, so that $y_2 = R \hat{e}$. We consider the limit as $R \to \infty$.  

The Hamiltonian of this system can be written as
\begin{align}
    H(R) &= -\frac{1}{2m_e} \Delta_{x_1} - \frac{1}{2m_e} \Delta_{x_2} + \frac{e^2}{\lvert{x_1 - x_2}\rvert} - \frac{e^2 Z}{\lvert{x_1}\rvert} - \frac{e^2 Z}{\lvert{x_2}\rvert} - \frac{e^2 Z}{\lvert{x_1 - R \hat{e}}\rvert} - \frac{e^2 Z}{\lvert{x_2 - R \hat{e}}\rvert}.
\end{align}
Consider, without loss of generality, the nuclear partition $\mathcal{N} = \{\{1\}, \{2\}\}$ and the electron partitions $\mathcal{E} = \{\{1\}, \{2\}\}$ and $\mathcal{E}' = \{\{2\}, \{1\}\}$. Then, for the cluster decomposition $\mathcal{N} \mathcal{E}$, we have
\begin{equation}
    H_{\mathcal{E}_1} = -\frac{1}{2m_e} \Delta_{x_1} - \frac{e^2 Z}{\lvert{x_1}\rvert}, \qquad H_{\mathcal{E}_2} = -\frac{1}{2m_e} \Delta_{x_2} - \frac{e^2 Z}{\lvert{x_2 - R \hat{e}}\rvert},
\end{equation}
and
\begin{align}
    I_{\mathcal{E}} = \frac{e^2}{\lvert{x_1 - x_2}\rvert} - \frac{e^2 Z}{\lvert{x_2}\rvert} - \frac{e^2 Z}{\lvert{x_1 - R \hat{e}}\rvert},
\end{align}
and analogously for the decomposition $\mathcal{N} \mathcal{E}'$. Notice that the clusters are just single hydrogen atoms with interaction $I_{\mathcal{E}}$ between them. Let $E_0$ denote the ground state energy of a neutral hydrogen atom and let $\psi_i(x)$ for $i =1,2$ denote the normalized ground state of each hydrogen atom, localized at $y_i$. Then, 
\begin{align}
    \langle \psi_1 \otimes \psi_2, H(R) \psi_1 \otimes \psi_2 \rangle &= 2 E_0 + o(1), \label{H(R)_sigma1}\\
    \langle \psi_2 \otimes \psi_1, H(R) \psi_2 \otimes \psi_1 \rangle &= 2 E_0 + o(1), \label{H(R)_sigma2}
\end{align}
where in (\ref{H(R)_sigma1}) and (\ref{H(R)_sigma2}) we used that $\langle \psi_1 \otimes \psi_2, I_{\mathcal{E}} \psi_1 \otimes \psi_2 \rangle = o(1)$ in $R$ by the exponential localization of eigenfunctions (\ref{exponentiallocalizationofeigenfunctions}). It is then easy to see that, without symmetry, the ground state of the decomposed system $H_{\mathcal{E}}$ is $\psi_1 \otimes \psi_2$ and similarly $\psi_2 \otimes \psi_1$ is for $H_{\mathcal{E}'}$. Thus the lowest eigenvalue of $H_\infty$ must have a double degeneracy and so the gap condition cannot hold. However, symmetrizing the ground states yields $S(\psi_1 \otimes \psi_2) = S(\psi_2 \otimes \psi_1)$, eliminating the degeneracy of $H_\infty$. 
\end{example}

\subsection{Re-statement of main result and proof} \label{mainresult_proof_subsection}

We now re-state our main result as Theorem \ref{GapConditionholdsTheorem} and afterwards give the proof. In the proof we defer the justification of two statements to Theorem \ref{clusterlimittheorem_symmetric}.

\begin{theorem}\label{GapConditionholdsTheorem}
(Global gap condition for bosons) Let $M \geqslant 2$ and Assumption \ref{localneutrality_assumption} hold. Then, there exists $\delta > 0$ such that for all $y \in \R^{3M}$, 
\begin{equation}
    E_1^S(y) - E_0^S(y) \geqslant \delta > 0, 
\end{equation}
where $E_0^S(y) = \inf \sigma(H_{N,M}^S(y))$ and $E_1^S(y) = \inf \left(\sigma(H_{N,M}^S(y)) \setminus \{E_0^S(y)\} \right)$.
\end{theorem}

\begin{remark}
With very strong non-degeneracy assumptions on ground states of the system and any cluster decompositions, we can also show that there exists $\delta > 0$ such that for all $y \in \R^{3M}$, 
\begin{equation}
    E_1^A(y) - E_0^A(y) \geqslant \delta > 0, \quad \delta \text{ independent of } y,
\end{equation}
where $E_1^A(y) = \inf \left(\sigma(H_{N,M}^A(y)) \setminus \{E_0^A(y)\} \right)$ in the fermionic case. To prove this, one can repeat the proof of Theorem \ref{GapConditionholdsTheorem} by replacing the subspace $\mathcal{H}_s$ with $\mathcal{H}_a$ and the projection $S$ with $A$.
\end{remark}

\begin{proof}[Proof of Theorem \ref{GapConditionholdsTheorem}]
In what follows, we drop the superscript $S$ denoting the symmetric case. Following Lemma \ref{Rto0limitlemma} b), it is sufficient to examine the spectral gap for the electronic Hamiltonian $H_{N,M}(y)$ without $V_n(y)$, i.e. without the nucleus-nucleus interaction terms. Hence, we replace $H_{N,M}(y)$ with $\tilde{H}_{N,M}(y)$, where
\begin{equation} \label{Hbo(y)_without_Vn}
    \tilde{H}_{N,M}(y) = - \sum_{j=1}^N \frac{1}{2m_e} \Delta_{x_j} + \sum_{i=1}^{N-1} \sum_{j=1}^N \frac{e^2}{\lvert{x_i - x_j}\rvert} + \frac{1}{2} \sum_{i=1}^N \sum_{j=1}^M \frac{e^2 Z_j}{\lvert{x_i - y_j}\rvert}.
\end{equation}
Of course, Lemma \ref{admissiblepartitionlemma} applies to $\tilde{H}_{N,M}(y)$ as well, and we drop the tilde when we apply the cluster breakup of $\tilde{H}_{N,M}$, i.e. (cf. \eqref{Hbo=Hcluster+Interaction}) 
\begin{equation} 
    \tilde{H}_{N,M}(y^n) = H_{\mathcal{E}}(Y^n) + I_{\mathcal{E}}(y^n),
\end{equation}
where $H_{\mathcal{E}}(Y^n)$ and $I_{\mathcal{E}}(y^n)$ are given by \eqref{Hboclusters}.

Aiming to argue by contradiction, assume there exists a sequence $\{y^n\}_{n =1}^\infty$ such that 
\begin{equation} \label{gapconditioncontradiction}
    E_1(y^n) - E_0(y^n) \to 0.
\end{equation}
After translating so that $y_1^n = 0$, if $\max_{i \neq j} \{ \nrm{y_i^n - y_j^n} \}$ is a bounded sequence, then it must have a convergent subsequence, so $y^{n_k} \to y^0$. By the continuity of $E(y)$ (see Section \ref{SE_operator_subsection}), this implies that $E(y^0) = E_1(y^0)$, contradicting the non-degeneracy of the ground state. Hence, we may assume  $\max_{i \neq j} \{ \nrm{y_i^n - y_j^n} \} \to \infty$. Then, using Lemma \ref{admissiblepartitionlemma}, we obtain an admissible partition $\mathcal{N}$, which we now consider fixed.

The partitions $\mathcal{E} \in \mathcal{A}_\mathcal{N}$ may assign electrons in such a way that some clusters are neutral and some are ions. Let us examine which configurations could yield suitable limits for the system described by $\tilde{H}_{N,M}(y^n)$, in its ground state or first excited state. 

In the case of neutral clusters, we denote  
\begin{equation}
    E_{j, 0} = \inf \sigma (H_{\mathcal{E}_j}), \qquad E_{j,1} = \inf \left(\sigma (H_{\mathcal{E}_j}) \setminus \{E_{j,0}\} \right).
\end{equation}
Similarly in the case of ions, we denote 
\begin{equation}
    E_{j,0}^{\delta_j} = \inf \sigma (H_{\mathcal{E}_j}), \qquad E_{j,1}^{\delta_j} = \inf \left( \sigma (H_{\mathcal{E}_j}) \setminus \{E_{j,0}^{\delta_j}\} \right),
\end{equation}
where $\delta_j$ is the ionic charge 
\begin{equation}
    \delta_j \equiv \nrm{\mathcal{E}_j} - \tilde{Z}_j,
\end{equation}
and recall $\tilde{Z}_j$ is the total nuclear charge, $ \tilde{Z}_j = \sum_{i \in \mathcal{N}_j} Z_i$ (note the cluster is neutral if $\delta_j = 0$ or $\nrm{\mathcal{E}_j} = \tilde{Z}_j$). The eigenvalues of the $H_{\mathcal{E}}$ Hamiltonian will be sums over all clusters of the eigenvalues of the clusters. 

For $\delta_j \geqslant 0$ (positive or neutral ions), $E_{j,0}^{\delta_j}$ is a simple eigenvalue below the continuum, and $\sigma(H_{\mathcal{E}_j})$ has an infinite amount of eigenvalues below its essential spectrum (see Theorem \ref{Hbospectrumcontinuous}). Hence for positive/neutral ions, 
\begin{equation} \label{Ejeigenstate}
    E_{j,0} < E_{j,1} < E_{j,2} < \inf \sigma_{ess} (H_{\mathcal{E}_j, 1}) = E_{j,0}^{1} < E_{j,0}^2 < \dots < E_{j,0}^{\tilde{Z}_j - 1}.
\end{equation}

If $\delta_j < 0$ is sufficiently negative (i.e. if a cluster of nuclei gains too many electrons), it may be the case that $E_{j,0}^{\delta_j}$ is not an eigenvalue below the continuum, i.e. $\sigma (H_{\mathcal{E}_j, \delta_j}) = [E_{j,0}^{\delta_j}, \infty)$. For such a non-binding result for molecules we refer back to \eqref{Lieb_nonbinding} and the discussion surrounding.
% For a molecule of $\nrm{\mathcal{N}_j}$ nuclei and total charge $\tilde{Z}_j$, the maximum number of electrons $N_{c_j}$ such that $H_{\mathcal{E}_j, N_{c_j} - \tilde{Z}_j}$ has a well-defined ground state satisfies $N_{c_j}  < 2 \tilde{Z}_j + \nrm{\mathcal{N}_j}$, or equivalently
% \begin{equation}
%     N_{c_j} - \tilde{Z}_j < \tilde{Z}_j + \nrm{\mathcal{N}_j}.
% \end{equation}
% This yields a bound on the maximum negative ionization. 
Hence, if $0 > -(\tilde{Z}_j + \nrm{\mathcal{N}_j}) \geqslant \delta_j$, then the negative ion with charge $\delta_j$ has no well-defined ground state. By the HVZ Theorem (Theorem \ref{Hbospectrumcontinuous}), this implies
\begin{equation} \label{Ej0noteigenstate}
    E_{j,0}^{-(\tilde{Z}_j + \nrm{\mathcal{N}_j})} = E_{j,0}^{-(\tilde{Z}_j + \nrm{\mathcal{N}_j}) - 1} = \dots = E_{j,0}^{-(\tilde{Z}_j + \nrm{\mathcal{N}_j}) - l} = \dots 
\end{equation}
for all integers $l > 0$.

Otherwise (if $0 > \delta_j > -(\tilde{Z}_j + \nrm{\mathcal{N}_j})$), we can only say, (with the exception of the non-existence of \ce{H^--}, see \cite{EL-1}),
\begin{equation} \label{Ejeigenstate_negative_deltaj}
    E_{j,0}^{-(\tilde{Z}_j + \nrm{\mathcal{N}_j}) + 1} \leqslant E_{j,0}^{-(\tilde{Z}_j + \nrm{\mathcal{N}_j}) + 2}\leqslant \dots \leqslant E_{j,0}.
\end{equation}

Combining (\ref{Ej0noteigenstate}), (\ref{Ejeigenstate}), and (\ref{Ejeigenstate_negative_deltaj}) we obtain the following chain of inequalities relating the ground state energies (and in the neutral case, also the first excited state energy) of the ions of a molecule of total charge $\tilde{Z}_j$ and number of nuclei $\nrm{\mathcal{N}_j}$. For any integer $l > 0$,
\begin{align}
     \dots &= E_{j,0}^{-(\tilde{Z}_j + \nrm{\mathcal{N}_j}) - l} = \dots = E_{j,0}^{-(\tilde{Z}_j + \nrm{\mathcal{N}_j}) - 1} = E_{j,0}^{-(\tilde{Z}_j + \nrm{\mathcal{N}_j})} \leqslant E_{j,0}^{-(\tilde{Z}_j + \nrm{\mathcal{N}_j}) + 1} \nonumber \\
     &\leqslant \dots \leqslant E_{j,0}^{-1} \leqslant E_{j,0} < E_{j,1} < E_{j,2} < E_{j,0}^1 < E_{j,0}^2 < \dots < E_{j,1}^{\tilde{Z}_j - 1}. \label{eigenvaluesofClusters}
\end{align}
 
By Assumption \ref{localneutrality_assumption} the configuration of lowest energy is the partition of the system into neutral clusters, each in their ground states. We denote this energy by $E_{\infty, 0}$, or more precisely,
\begin{equation} \label{E_{infty_0}_def}
    E_{\infty, 0}(Y^n) = \sum_{j=1}^k E_{j, 0}(Y^n_{\mathcal{N}_j}),
\end{equation}
so we expect the limit 
\begin{equation} \label{groundstatedecomplimit_notTheorem}
    \lim_{n \to \infty} E_0(y^n) = E_{\infty,0}(Y^0).
\end{equation}
For the next energy value, our candidates must be neutral clusters with one cluster in the first excited state (and the rest in ground states), which has energy
\begin{equation} \label{E_{infty,1}_def1}
    E_{\infty,0}^1(Y^n) = \min_{1 \leqslant l \leqslant k} \left\{\sum_{\substack{j=1 \\ j \neq l}}^k E_{j,0}(Y^n_{\mathcal{N}_j}) + E_{l, 1}(Y^n_{\mathcal{N}_l}) \right\},
\end{equation}
or some configuration of ions in their ground states, which has energy
\begin{equation} \label{E_{infty,1}_def2}
     E_{\infty,0}^I(Y^n) = \min_{I = (\delta_1, \dots, \delta_k)} \left\{\sum_{j=1}^k E_{j, 0}^{\delta_j}(Y^n_{\mathcal{N}_j}) \right\},
\end{equation}
where $I = (\delta_1, \dots, \delta_k)$ is given by (\ref{ioncharges}), and each $E_{j,0}^{\delta_j}$ is an eigenvalue below the continuum. In this case we expect 
\begin{equation} \label{1stexcitedstatelimit_notTheorem}
    \lim_{n \to \infty} E_1(y^n) = \min \left\{ E_{\infty,0}^1(Y^0), \:\: E^I_{\infty,0}(Y^0) \right\}.
\end{equation}

\begin{lemma} \label{eigenstate_ionicclusters_lemma}
The minimizer of the right-hand side of \eqref{1stexcitedstatelimit_notTheorem} must have a corresponding eigenstate.
\end{lemma}

\begin{proof}[Proof of Lemma \ref{eigenstate_ionicclusters_lemma}] (See also Lemma $6$ in \cite{lewin:hal-00093523}.) In the case of neutral clusters with one cluster in an excited state, \eqref{E_{infty,1}_def1}, this follows using Theorem \ref{Hbospectrumcontinuous} b). The case of ions, \eqref{E_{infty,1}_def2}, requires a proof. Suppose that 
\begin{equation} \label{Ek0Deltaj<Ek0Deltaj+1_0}
     \min_{\substack{1 \leqslant l \leqslant k \\ I = (\delta_1, \dots, \delta_k)}} \left\{ E_{\infty,0}^1(Y^0), \:\: E^I_{\infty,0}(Y^0) \right\} = \sum_{j=1}^k E_{j, 0}^{\delta_j}(Y^n_{\mathcal{N}_j}),
\end{equation}
for some configuration of ions $I = (\delta_1, \dots, \delta_k)$ given by (\ref{ioncharges}), and in what follows we drop the $Y^n_{\mathcal{N}_j}$ dependence for notational convenience. We construct an eigenstate corresponding to the right hand side of \eqref{Ek0Deltaj<Ek0Deltaj+1_0} by finding eigenstates for each energy $E_{j, 0}^{\delta_j}$ and taking the symmetrized tensor product over $1 \leqslant j \leqslant k$.

In view of the HVZ theorem, it is enough to show that $E_{i,0}^{\delta_i} < E_{i,0}^{\delta_i + 1}$ for all $1 \leqslant i \leqslant k$. Assume that for some $j$ we have $E_{j,0}^{\delta_j} = E_{j,0}^{\delta_j + 1}$. Then, by the HVZ and Zhislin theorems, this can occur only if $\delta_j < 0$. Accordingly, there must exist an $i$ such that $\delta_i > 0$, and as a consequence $E_{i,0}^{\delta_i} < E_{i,0}^{\delta_i + 1}$. It follows that
\begin{align}
    \sum_{l=0}^k E_{l,0}^{\delta_l} &= E_{j,0}^{\delta_j} + \sum_{\substack{l=0 \\ l \neq i}}^k E_{l,0}^{\delta_l} > E_{j,0}^{\delta_j} + E_{i,0}^{\delta_i + 1} + \sum_{\substack{l=0 \\ l \neq i, j}}^k E_{l,0}^{\delta_l},
\end{align}
contradicting the fact that $I = (\delta_1, \dots, \delta_k)$ is an ionic minimizer. Hence every ionic minimizer of \eqref{Ek0Deltaj<Ek0Deltaj+1_0} must have a corresponding ground state.
\end{proof}

We defer the proofs of \eqref{groundstatedecomplimit_notTheorem} and \eqref{1stexcitedstatelimit_notTheorem} to Theorem \ref{clusterlimittheorem_symmetric}. Assuming their validity, the proof of Theorem \ref{GapConditionholdsTheorem} now concludes. Indeed, using Assumption \ref{localneutrality_assumption}, 
\begin{equation} \label{localneutralityforclusters}
  E_{\infty,0}^I > E_{\infty,0},
\end{equation}
where $I = (\delta_1, \dots, \delta_k)$ given by (\ref{ioncharges}). As $E_{l,1} > E_{l,0}$ for all $1 \leqslant l \leqslant k$, there exists a constant $\delta > 0$ such that $E_{\infty, 1} \geqslant E_{\infty, 0} + \delta$ (note in the case of part b) we have assumed that all ground state energies are non-degenerate). Hence in all cases it is clear to see that $\lim_{n \to \infty} E_1(y^n) - E_0(y^n) > 0$, so we obtain a contradiction. 
\end{proof}

%%%%%%%%%%%%%%%%%%%%%%%%%%%%%%%%%%%%%%%%%%%%%%%%%%%%%%%%%%%%%%%%%%%%%%%%%%%
\section{Eigenvalue limits} \label{eigenvaluelimits_section}
%%%%%%%%%%%%%%%%%%%%%%%%%%%%%%%%%%%%%%%%%%%%%%%%%%%%%%%%%%%%%%%%%%%%%%%%%%%

To conclude the proof of Theorem \ref{GapConditionholdsTheorem}, it remains to justify \eqref{groundstatedecomplimit_notTheorem} and \eqref{1stexcitedstatelimit_notTheorem}. We summarize these statements in the following theorem, which is technical. 

\begin{theorem} \label{clusterlimittheorem_symmetric}
Let Assumption \ref{localneutrality_assumption} hold. Consider a sequence $\{y^n\}_{n=1}^\infty$ such that $\lvert{y^n_i - y^n_j}\rvert \to \infty$ for at least one pair of coordinates $i \neq j$. By Lemma \ref{admissiblepartitionlemma}, an admissible partition $\mathcal{N}$ exists for the Hamiltonian $\tilde{H}_{N,M}(y)$, and along the admissible partition sequence the following limits converge to the following values.
\begin{align}
\lim_{n \to \infty} E_0(y^n) &= E_{\infty,0}(Y^0) = \sum_{j=1}^k E_{j, 0}(Y^0_{\mathcal{N}_j}), \label{groundstatedecomplimit} \\
\lim_{n \to \infty} E_1(y^n) &= E_{\infty, 1}(Y^0) = \min_{\substack{1 \leqslant l \leqslant k \\ I = (\delta_1, \dots, \delta_k)}} \left\{ E_{\infty,0}^l(Y^0), \:\: E^I_{\infty,0}(Y^0) \right\}, \label{1stexcitedstatelimit}
\end{align}
where
\begin{align}
    E_{\infty,0}^l(Y^n) &= \sum_{\substack{j=1 \\ j \neq l}}^k E_{j,0}(Y^n_{\mathcal{N}_j}) + E_{l, 1}(Y^n_{\mathcal{N}_l}), \quad 
    E_{\infty,0}^I(Y^n) = \sum_{j=1}^k E_{j, 0}^{\delta_j}(Y^n_{\mathcal{N}_j}),
\end{align}
and 
\begin{equation} \label{Ioncharges_def2}
    I = \left\{\nrm{\mathcal{E}_j} - \sum_{i \in \mathcal{N}_j} Z_i \right\}_{j=1}^k.
\end{equation}
Recall that here we are considering the completely symmetric case (bosonic statistics), dropping the $S$ dependence in the notation.
\end{theorem}

We now provide a direct proof of Theorem \ref{clusterlimittheorem_symmetric} using the time-independent FS map.

\subsection{Proof of Theorem \ref{clusterlimittheorem_symmetric}} \label{clusterlimittheorem_proof_subsection}
We break the proof into $3$ main steps.

\textbf{Step 1: }\textit{Defining the FS map projection.}  We denote the unique ground state eigenfunction of $H_{\mathcal{E}_j}(Y^n_{\mathcal{N}_j})$ by $\psi_{\mathcal{E}_j, 0}^{(n)} = \psi_{\mathcal{E}_j, 0}^{(n)}(\tilde x_{\mathcal{E}_j})$ (recall the non-degeneracy of bosonic ground states) and the first excited state by $\psi_{\mathcal{E}_j,1,p}^{(n)} = \psi_{\mathcal{E}_j,1,p}^{(n)}(\tilde x_{\mathcal{E}_j})$, ordered by multiplicity $p = 1, \dots, m_j$. Recall that these eigenstates will always exist, as they correspond to neutral clusters. We define
\begin{equation} \label{neutralcluster_eigenstates}
    \psi_0^{\mathcal{E}, n}(x) = \tau_n^{-1} \bigotimes_{j=1}^k \psi_{\mathcal{E}_j,0}^{(n)}(\tilde x_{\mathcal{E}_j}), \qquad \psi_{l,p}^{\mathcal{E}, n}(x) = \tau_n^{-1} \bigotimes_{\substack{j=1 \\ j \neq l}}^k \psi_{\mathcal{E}_j,0}^{(n)}(\tilde x_{\mathcal{E}_j}) \otimes \psi_{\mathcal{E}_l,1,p}^{(n)}(\tilde x_{\mathcal{E}_j}),
\end{equation}
which are eigenfunctions of $H_{\mathcal{E}}(Y^n)$ corresponding to the eigenvalues $E_{\infty, 0}(Y^n)$ and $E_{\infty}^l(Y^n)$, respectively. Recall the translation $\tau_n$ is given in \eqref{clustertranslation} ad and the $\tilde x$ are shifted coordinates, i.e. $\phi(\tilde x) = \tau_n \phi(x)$ (cf. Lemma \ref{admissiblepartitionlemma}).

% The latter are among the candidates for the minimization problem on the right-hand side of \eqref{1stexcitedstatelimit}.  

We refer to minimizers of ionic clusters in their ground states, cf. \eqref{1stexcitedstatelimit}, as \textit{ionic minimizers}. These ground states are of the form 
\begin{equation} \label{ionic_minimizer_groundstate}
    \psi_I^{\mathcal{E}, n}(x) = \tau_n^{-1} \bigotimes_{j=1}^k \psi_{\mathcal{E}_j,0}^{(n), \delta_j}(\tilde x_{\mathcal{E}_j}),
\end{equation}
where $\psi_{\mathcal{E}_j,0}^{(n), \delta_j} = \psi_{\mathcal{E}_j,0}^{(n), \delta_j}(\tilde x_{\mathcal{E}_j})$ is the ground state corresponding to ion ground state energy $E_{j,0}^{\delta_j}(Y^n_{\mathcal{N}_j})$ of the ionic cluster $H_{\mathcal{E}_j}(Y^n_{\mathcal{N}_j})$ with ionic charge $\delta_j = \nrm{\mathcal{E}_j} - \tilde{Z}_j$ (cf. \ref{Ioncharges_def2}). By Lemma \ref{eigenstate_ionicclusters_lemma} any ionic minimizer of \eqref{1stexcitedstatelimit} must have a ground state and hence \eqref{ionic_minimizer_groundstate} exists. By construction we have $H_{\mathcal{E}}(Y^n) \psi_I^{\mathcal{E}, n} = E_{\infty, 0}^I(Y^n) \psi_I^{\mathcal{E}, n}$.  

Note that given a partition $\mathcal{E}'$ of the electrons, for any permutation $\pi \in S_N$ there exists another partition $\mathcal{E} = \mathcal{E}(\pi)$ such that $T_\pi \psi_{j}^{\mathcal{E}',n} = \psi_j^{\mathcal{E}(\pi), n}$ and $T_\pi \psi_I^{\mathcal{E}', n} = \psi_I^{\mathcal{E}(\pi), n}$ (we drop the dependence of $a(\pi)$ on $\mathcal{E}'$ to emphasize only the final position of the permuted and partitioned coordinates). 

Let $\tilde{S}(\psi) := \frac{S(\psi)}{\lVert{S(\psi)}\rVert}$ denote the normalized symmetrizer and define the projection $P = P(y^n)$ to be the orthogonal projection of finite rank onto the subspace $\textrm{span }\mathcal{F}$, where $\mathcal{F}$ consists of the collection of normalized symmetrized ground states of neutral clusters in their ground states, neutral clusters in ground states with one cluster in the first excited state, and ionic minimizers. More precisely, $\mathcal{F}$ consists of the collection of states \eqref{neutralcluster_eigenstates} and \eqref{ionic_minimizer_groundstate} symmetrized and normalized, i.e.
\begin{align}
    &\mathcal{F} = \left\{ \tilde{S}(\psi_0^{\mathcal{E}, n}) \right\} \cup \left\{ \tilde{S}(\psi_{l,p}^{\mathcal{E}, n}) \right\} \cup \left\{ \tilde{S}(\psi_I^{\mathcal{E}, n}) \right\}.
    \label{defof_mathcalF} 
\end{align}

Note that $P$ and $\mathcal{F}$ depend on $n$, even though we suppress this in the notation. To compute $P$, we use the fact that the elements of $\mathcal{F}$ are almost pairwise orthogonal for large enough $n$ (which we defer to Lemma \ref{FSMap_estimates} and in particular estimate \eqref{almostorthogonal_estimate}). This implies that the matrix $(g_{ij}^n) = (\langle \phi_i^n, \phi_j^n \rangle) = I + o(1)$, where $\phi_i^n, \phi_j^n \in \mathcal{F}$. Then, we can write
\begin{equation} \label{P_almostorthogonal_braket}
    P = \sum_{\phi_i^n, \phi_j^n \in \mathcal{F}} |\phi_i^n \rangle g^{n, ij} \langle \phi_j^n |,
\end{equation}
where $(g^{n,ij})$ is the inverse of the matrix $(g_{ij}^n)$. Then we also have that $(g^{n, ij}) =I + o(1)$, and hence
\begin{equation} 
    P = \sum_{\phi_i^n \in \mathcal{F}} |\phi_i^n \rangle \langle \phi_i^n | + o(1),
\end{equation}
or explicitly,
\begin{align} 
    P \phi &= \tilde{S}(\psi_0^{\mathcal{E}, n}) \langle \tilde{S}(\psi_0^{\mathcal{E}, n}), \phi \rangle + \sum_{l=1}^k \sum_{p = 1}^{m_l} \tilde{S}(\psi_{l, p}^{\mathcal{E}, n}) \langle \tilde{S}(\psi_{l, p}^{\mathcal{E}, n}), \phi \rangle + \sum_{I \:\:\text{i.m.}} \tilde{S}(\psi^{\mathcal{E},n}_I) \langle \tilde{S}(\psi^{\mathcal{E},n}_I), \phi \rangle + o(1), \label{P_almostorthogonal_finalexpression}
\end{align}
where $\mathcal{E}$ is some partition of the electrons into neutral clusters (note that the function $S(\psi^{\mathcal{E}})$ is independent of the choice of $\mathcal{E}$). 

In what follows we drop the $o(1)$ term in \eqref{P_almostorthogonal_finalexpression} as in all cases it will be clear that its contribution is negligible. Define also the orthogonal complement by $P^\perp(y^n) = 1 - P(y^n)$ and let $\tilde{H}_{N,M}^\perp(y^n) = P^\perp(y^n) \tilde{H}_{N,M}(y^n) P^\perp(y^n)$. \\

\textbf{Step 2: }\textit{Verifying that the FS map is well-defined.} In order to prove the existence of the FS map (cf. Theorem \ref{FSmapTheorem}) we first remark that $\textrm{Ran }P(y^n) \subset \mathcal{D}(\tilde{H}_{N,M})$, since the range of $P(y^n)$ is spanned by the elements of $\mathcal{F}$, all of which belong to $H^2(\mathbb{R}^{3N})$. Additionally, by \eqref{P_almostorthogonal_braket}, we have that 
\begin{equation}
    \tilde{H}_{N,M}(y^n) P(y^n) = \sum_{\phi_i^n, \phi_j^n \in \mathcal{F}} |\tilde{H}_{N,M}(y^n) \phi_i^n \rangle g^{n, ij} \langle \phi_j^n |,
\end{equation}
and since each $\phi_i$ is in $H^2(\mathbb{R}^{3N})$ we have that $\lVert{\tilde{H}_{N,M}(y^n) P(y^n)}\rVert < \infty$. It remains to show that for sufficiently large $n$, there exists a constant $G > 0$ such that 
\begin{equation} \label{FSmap_condition}
    \tilde{H}_{N,M}^\perp(y^n) - E_1(y^n) \geqslant G. 
\end{equation}
This implies that the resolvent $(\tilde{H}_{N,M}(y^n)^\perp - \lambda)^{-1}$ is well-defined for all $\lambda \leqslant E_1(y^n)$, and we have the estimate
\begin{equation} \label{Hboresolvent_E0(yn)_estimate}
    \lVert{(\tilde{H}_{N,M}^\perp(y^n) - \lambda)^{-1}}\rVert \leqslant \frac{1}{G}, \quad \forall \: \lambda \leqslant E_1(y^n).
\end{equation}
In order to prove (\ref{FSmap_condition}) we make use of the IMS localization formula (the proof of existence is standard, see for example \cite{IMS-1} or \cite{GS}). Denote by $\mathcal{B}_\mathcal{N}$ the set of all partitions $b = \{\mathcal{E}_1, \dots, \mathcal{E}_{k+1}\}$ of $\{1, \dots, N\}$, assigning electrons to one of the $k$ clusters $\mathcal{E}_j$ for $1 \leqslant j \leqslant k$, or away from any cluster in the case of $\mathcal{E}_{k+1}$. 

\begin{lemma}\label{IMS_PO1_lemma}
Given an admissible partition of the nuclei, there exists $R_n$ such that $R_n \geqslant 1$, $R_n \to \infty$ as $n \to \infty$,
and
\begin{equation} \label{R=R(n)choice}
    2 R_n (C_i + C_j) \leqslant \frac{r_{ij}^n}{2}
\end{equation}
for all $i \neq j$ and each $n$ sufficiently large. Here, recall the definitions of the cluster size $C_j$ and inter-cluster radius $r_{ij}^n$ from Lemma \ref{admissiblepartitionlemma}. Then, there exists a family $\{J_{b,n}\}$ of $C^\infty(\mathbb{R}^{3N}, [0,1])$ functions such that, for any partitions $b$,
\begin{align}
    &\partial^\beta J_{b,n} \in L^\infty \textrm{ for } 0 \leqslant \lvert{\beta}\rvert \leqslant 2, \label{JaR_bounded} \\
    &\sum_{b \in \mathcal{B}_\mathcal{N}} (J_{b,n})^2 = 1, \label{JaR_sum} \\
    &\lvert{\nabla J_{b,n}}\rvert^2 \leqslant \frac{d}{R_n}  \textrm{ for some constant } d > 0. \label{JaR_nablato0}
\end{align}
$J_{b,n}$ represents the localization of the $l^{th}$ electron in either $B(z_j^n, 2 R_n C_j)$ for $1 \leqslant j \leqslant k$ containing the $j^{th}$ cluster or away from any cluster, according to the partition $b$; i.e. the one-electron density $\lvert{J_{b,n}(\dots, x_l, \dots)}\rvert = 0$ outside of these sets.

Furthermore, let $S_{j,b}$ denote the symmetrizer over only the coordinates with indices belonging to $\mathcal{E}_j$. Then, $S_{j,b} J_{b,n} = J_{b,n}$ and as a consequence, 
\begin{equation}
    [J_{b,n}, S_{j,b}] = 0. \label{JaR_symmetrizer}
\end{equation}
\end{lemma}

By the IMS localization formula, it follows that 
\begin{equation}
    \tilde{H}_{N,M}(y^n) = \sum_{b \in \mathcal{B}_\mathcal{N}} J_{b,n} \tilde{H}_{N,M}(y^n) J_{b,n} - \lvert{\nabla J_{b,n}}\rvert^2,
\end{equation}
and hence using (\ref{JaR_nablato0}),
\begin{equation} \label{Hperp-E_0_below0}
    \tilde{H}_{N,M}^\perp(y^n) \geqslant \sum_{b \in \mathcal{B}_\mathcal{N}} P^\perp J_{b,n} \tilde{H}_{N,M}(y^n) J_{b,n} P^\perp - \frac{\tilde{d}}{R},
\end{equation}
for some constant $\tilde{d} > 0$.

\begin{proposition} \label{PperpJaHaJaPperp_belowestimates}
For all partitions $b \in \mathcal{B}_{\mathcal{N}}$, we have that 
\begin{align} \label{PperpJaHaJaPperp_below}
    &P^\perp J_{b,n} \tilde{H}_{N,M}(y^n) J_{b,n} P^\perp \geqslant \left(E_{\infty,1}(Y^n) + G - o(1) \right) P^\perp J_{b,n}^2 P^\perp - o(1), 
\end{align}
where recall $E_{\infty, 1}(Y^n)$ is defined in \eqref{1stexcitedstatelimit}, and $G > 0$ is a constant independent of $n$ such that
\begin{equation} \label{E0+C<0}
   E_{\infty,1}(Y^n) + G < 0. 
\end{equation}
\end{proposition}

Using (\ref{PperpJaHaJaPperp_below}) and (\ref{JaR_sum}), from (\ref{Hperp-E_0_below0}) we obtain
\begin{equation} \label{Hperp-E_0_below1}
    \tilde{H}_{N,M}^\perp(y^n) \geqslant \left(E_{\infty,1} + G - o(1) \right) P^\perp - o(1).
\end{equation}
Using (\ref{E0+C<0}) and $P > 0$ we have that
\begin{equation}
    \left(E_{\infty,1} + G - o(1) \right) P^\perp \geqslant E_{\infty,1} + G - o(1). \label{Hperp-E_0_below2}
\end{equation}
Next, we estimate $E_1$ in terms of $E_{\infty, 1}$ by using the variational principle. 

\begin{lemma} \label{FSMap_estimates}
In the limit as $n \to \infty$, we have the following estimates for the variational principle. For any $\phi_i^n \in \mathcal{F}$ (where recall $\mathcal{F}$ is defined in (\ref{defof_mathcalF})) denote by $\lambda_i$ its corresponding eigenvalue, i.e. $H_{\mathcal{E}} \phi_i^n = \lambda_i \phi_i^n$. Then, for all $\phi_i^n, \phi_j^n \in \mathcal{F}$, we have
\begin{align}
    &\langle \phi_i^n, \phi_j^n \rangle = o(1) \delta_{ij}, \label{almostorthogonal_estimate} \\
    &\langle \phi_i, \tilde{H}_{N,M}(y^n) \phi_j \rangle = \lambda_i \delta_{ij} + o(1), \label{SHboS_estimate}  \\
    &\langle \phi_i, \tilde{H}_{N,M}(y^n) P^\perp (\tilde{H}_{N,M}^\perp(y^n) - E(y^n))^{-1} P^\perp \tilde{H}_{N,M}(y^n) \phi_j \rangle = o(1), \label{FSresolventterm_estimate}
\end{align}
where $\delta_{ij}$ denotes the Kronecker delta. 
\end{lemma}

By the min-max theorem for eigenvalues,
\begin{equation} \label{E1minmax}
    E_1(y^n) = \inf_{\substack{A \subset \mathcal{H}_s \\ \textrm{dim }A = 2}} \sup_{\substack{\psi \in A \\ \lVert{A}\rVert = 1}} \langle \psi, \tilde{H}_{N,M}(y^n) \psi \rangle.
\end{equation}
Choosing subspaces $A = \textrm{span} \{\tilde{S}(\psi_0^{\mathcal{E}, n}), \phi\}$ where $\phi \in \mathcal{F} \setminus \{\tilde{S}(\psi_0^{\mathcal{E}, n})\}$, we estimate from (\ref{E1minmax}) using (\ref{SHboS_estimate}) and $E_{\infty, 0} < E_{\infty, 1}$ to obtain
\begin{equation} \label{Hperp-E_0_below3}
    E_1(y^n) \leqslant E_{\infty, 1}(Y^n) + o(1).
\end{equation}
Combining (\ref{Hperp-E_0_below1}), (\ref{Hperp-E_0_below2}), and (\ref{Hperp-E_0_below3}) we obtain
\begin{equation}
    \tilde{H}_{N,M}^\perp(y^n) \geqslant E_1(y^n) + G - o(1),
\end{equation}
from which (\ref{FSmap_condition}) follows for $n$ sufficiently large and hence the FS map is well-defined. \\

\textbf{Step 3: }\textit{Using the properties of the FS map to estimate the eigenvalues.} By Corollary \ref{FScorollary}, the operator 
% \begin{align}
%     F_P(E(y^n)) &= P \tilde{H}_{N,M}(y^n) P \nonumber \\
%     &\hspace{15pt} - P \tilde{H}_{N,M}(y^n) P^\perp \left(\tilde{H}_{N,M}^\perp(y^n) - E(y^n) \right)^{-1} P^\perp \tilde{H}_{N,M}(y^n) P \label{FSmapHbo}
% \end{align}
\begin{align}
    F_P(\lambda) &= P \tilde{H}_{N,M} P - P \tilde{H}_{N,M} P^\perp \left(\tilde{H}_{N,M}^\perp - \lambda \right)^{-1} P^\perp \tilde{H}_{N,M} P \label{FSmapHbo}
\end{align}
acting on $\textrm{Ran }P$ has a finite number of eigenvalues that we denote by $\nu_0(\lambda) \leqslant \nu_1(\lambda) \leqslant \dots \leqslant \nu_{\textrm{Rank }P}(\lambda)$ counting multiplicity, corresponding to the eigenvalues $E_0(y^n)$ and $E_1(y^n)$ via the fixed point problems
\begin{equation}
    E_0(y^n) = \nu_0(E_0(y^n)), \qquad E_1(y^n) = \nu_1(E_1(y^n)), 
\end{equation}
where recall that the ground state $E_0$ is non-degenerate. Hence, we use the min-max theorem to write 
\begin{align} 
    E_0(y^n) &= \inf_{\substack{\phi \in \textrm{ Ran }P \\ \lVert{\phi}\rVert = 1}} \langle \phi, F_P(E_0(y^n)) \phi \rangle = \langle \tilde{S}(\psi_0^{\mathcal{E}, n}), F_P(E_0(y^n)) \tilde{S}(\psi_0^{\mathcal{E}, n}) \rangle, \label{E_0(yn)expansion} 
\end{align}
and
\begin{align}
    E_1(y^n) &= \inf_{\substack{A \subset \textrm{ Ran }P \\ \textrm{dim }A = 2}} \sup_{\substack{\psi \in A \\ \lVert{A}\rVert = 1}} \langle \psi, F_P(E_1(y^n)) \psi \rangle. \label{E_1(yn)expansion}
\end{align}
We use the expansion (\ref{FSmapHbo}) and estimates (\ref{SHboS_estimate}) and (\ref{FSresolventterm_estimate}). Note that the infimum in (\ref{E_1(yn)expansion}) is taken over 2-dimensional subspaces of $\textrm{Ran }P$, as opposed to in (\ref{E1minmax}) where it was taken over 2-dimensional subspaces of the much larger space $\mathcal{H}_s$. Hence by repeating the argument preceding the estimate (\ref{Hperp-E_0_below3}) we improve it, obtaining
\begin{align} \label{E_0(yn)expansion1}
    E_0(y^n) =& E_{\infty, 0}(Y^n) + o(1), \\
    E_1(y^n) =& E_{\infty, 1}(Y^n) + o(1),
\end{align}
from which the limits (\ref{groundstatedecomplimit}) and (\ref{1stexcitedstatelimit}) follow and the proof of Theorem \ref{clusterlimittheorem_symmetric} concludes. 
\qed

% \subsection{Proof of Lemma \ref{IMS_PO1_lemma}} For $n$ large enough, $r^n_{ij} > 2(C_i + C_j)$ for $i \neq j$. Since $r^n_{ij} \to \infty$ as $n \to \infty$, one can choose a sequence $R_n$ such that $R_n \geqslant 1$, $R_n \to \infty$, (\ref{R=R(n)choice}) for each $n$, and such that the sets $B(z_j^n, 2 R_n C_j)$ are all disjoint. 

% Choose a function $g \in C^\infty(\mathbb{R}, [0,1])$ with $g(r) = 0$ for $r \leqslant 1$ and $g(r) = 1$ for $r \geqslant 2$. For 1 $\leqslant j \leqslant k$ define
% \begin{equation}
%     f_{j, R_n}(x) = 1 - g\left(\frac{\lvert{x - z_j^n}\rvert}{R_n C_j}\right),
% \end{equation}
% then
% \begin{equation}
%     \supp f_{j, R_n} \subset B(z_j^n, 2 R_n C_j), \quad f_{j, R_n} = 1 \textrm{ for } x \in B(z_j^n, R_n C_j).
% \end{equation}
% For $j = k+1$, define
% \begin{equation}
%     f_{k+1, R_n}(x) = 1 - \sum_{j=1}^k f_{j, R_n}. 
% \end{equation}
% For the partition $b \in \mathcal{B}_{\mathcal{N}}$ define
% \begin{equation}
%     h_{b, R_n}(x) = \prod_{j=1}^{k+1} \prod_{l \in \mathcal{E}_j} f_{j, R_n}(x_l),
% \end{equation}
% and then
% \begin{equation}
%     J_{b,n}(x) = \frac{h_{b, R_n}(x)}{(\sum_b h_{b, R_n}^2)^{1/2}}
% \end{equation}
% satisfies the required properties. In particular, note that in the denominator, $\sum_b h_{b, R_n}^2 \geqslant \frac{1}{4}$ as the sum consists locally of at most two terms and $a + b = 1$ implies $a^2 + b^2 \geqslant \frac{1}{4}$.
% \qed

\subsection{Proof of Proposition \ref{PperpJaHaJaPperp_belowestimates}} We first begin by defining the constant $G$. Let
\begin{equation}
    G = \min \left\{ G_1, G_2, G_3, G_4 \right\},
\end{equation}
where we define the positive constants $G_1, G_2, G_3,$ and $ G_4$ in the following ways. Then $G>0$ and we can see that (\ref{E0+C<0}) follows immediately as the ground state or first excited state energies of all clusters are negative.

First, we consider the smallest gap between the ground state energy and first excited state energy of neutral clusters,
\begin{equation}
    G_1 = \min_{1 \leqslant j \leqslant k} \left\{E_{j, 2} - E_{j, 1}, E_{j, 1} - E_{j, 0} \right\}.
\end{equation}
Clearly $G_1 > 0$ by Theorem \ref{Hbospectrumcontinuous} (cf. \eqref{eigenvaluesofClusters}). Next, we consider gaps between eigenvalues of the ionic minimizers, when they exist. 
\begin{equation} \label{G2_def}
    G_2 = \min_{\substack{1 \leqslant j \leqslant k, \\ I = (\delta_1, \dots, \delta_k) \text{ i.m. }}} \left\{E_{j,1}^{\delta_j} - E_{j,0}^{\delta_j} \right\},
\end{equation}
as long as the set is non-empty. $G_2$ is the smallest gap between the ground state energy and first excited state energy of ionic minimizers. $G_2 > 0$ follows directly from Lemma \ref{eigenstate_ionicclusters_lemma}. Next,
\begin{equation} \label{G3_def}
    G_3 = \min_{\substack{1 \leqslant j \leqslant k, \\ I = (\delta_1, \dots, \delta_k) \text{ i.m. } \\ \delta_j \geqslant 1 0}} \left\{E_{j,0}^{\delta_j} - E_{j,0}^{\delta_j -1} \right\},
\end{equation}
the smallest gap between the ground state energies of strictly positive ionic minimizers and the ionic minimizer gaining one electron (note this also includes the neutral energy, i.e. $E_{j,0}^{1} > E_{j,0}$). $G_3 > 0$ by Theorem \ref{Hbospectrumcontinuous} (cf. \eqref{eigenvaluesofClusters}), as we only consider strictly positive ions. Lastly, we define
\begin{equation} \label{G4_def}
    G_4 = \min_{\substack{I = (\delta_1, \dots, \delta_k) \\ \text{ not i.m.}}} \left\{E_{\infty, 0}^I - E_{\infty, 1} \right\},
\end{equation}
the smallest gap between the total ground state energy of non-minimizing ionic clusters and the total first excited state energy $E_{\infty, 1}$, which recall from \eqref{1stexcitedstatelimit} is the minimum over the total ground state energy of ionic clusters and neutral clusters in ground states except for one cluster in the first excited state.

Now, we move to showing the estimate \eqref{PperpJaHaJaPperp_below}. We consider the possibilities for the partition $b$ case by case. In what follows, let $\mathcal{E} = b \setminus \mathcal{E}_{k+1}$. If $\mathcal{E}_{k+1} = \emptyset$, then $\mathcal{E} \in \mathcal{A}_{\mathcal{N}}$. 

\subsubsection{Case I} $\lvert{\mathcal{E}_{k+1}}\rvert >0$. At least one electron is localized far away from any cluster. Then the clusters must contain at least one ion, and we can write
\begin{equation} \label{Hbo_case1}
    \tilde{H}_{N,M}(y^n) = H_{\mathcal{E}}(Y^n) + I_{\mathcal{E}}(Y^n) + H_{\mathcal{E}_{k+1}}(Y^n),
\end{equation}
where
\begin{equation}
    H_{\mathcal{E}_{k+1}}(Y^n) = \sum_{l \in \mathcal{E}_{k+1}} \left(-\frac{1}{2m_e} \Delta_{x_l} + I_l(Y^n) \right) + \sum_{\substack{i < j \\ i,j \in \mathcal{E}_{k+1}}} \frac{e^2}{\lvert{x_i - x_j}\rvert},
\end{equation}
and by $I_l(Y^n)$ we denote the interaction of the free electron $l$ with the clusters, i.e.
\begin{equation}
    I_l(Y^n) = \left(\tau_n^b\right)^{-1} \sum_{j=1}^k \left(\sum_{s \in \mathcal{E}_j} \frac{e^2}{\lvert{\tilde{x}_s - x_l + z_j^n}\rvert} - \sum_{s \in \mathcal{N}_j} \frac{e^2 Z_s}{\lvert{Y_s^n - x_l + z_j^n}\rvert} \right) \tau_n^b,
\end{equation}
with $\tau_n^b$ the translation subject to the partition $b$,
\begin{equation} \label{taun^b_def1}
    \tau_n^b \phi(x_1, \dots, x_N) = \phi(x_1 - \delta_1^{n, b}, \dots, x_N - \delta_N^{n, b}), 
\end{equation}
where now
\begin{equation} \label{taun^b_def2}
    \delta_i^{n, b} = 
    \begin{cases}
        z_j^n &\:\text{ for }\: i \in \mathcal{E}_j, \:1 \leqslant j \leqslant k, \\
        0 &\:\text{ for }\: i \in \mathcal{E}_{k+1}.
    \end{cases}
\end{equation}
$H_{\mathcal{E}}$ and $I_{\mathcal{E}}$ are given by (\ref{Hcluster}) and (\ref{clusterinteraction}), with $\tau_n$ replaced by $\tau_n^b$. We have the following estimates:
\begin{align}
    P^\perp J_{b,n} H_{\mathcal{E}} J_{b,n} P^\perp &\geqslant (E_{\infty, 1} + G) P^\perp J_{b,n}^2 P^\perp, \label{case1estimate1} \\
    P^\perp J_{b,n} I_{\mathcal{E}} J_{b,n} P^\perp &\geqslant -o(1) P^\perp J_{b,n}^2 P^\perp, \label{case1estimate2} \\
    P^\perp J_{b,n} H_{\mathcal{E}_{k+1}} J_{b,n} P^\perp &\geqslant -o(1) P^\perp J_{b,n}^2 P^\perp. \label{case1estimate3} 
\end{align}
Combining (\ref{Hbo_case1}) and (\ref{case1estimate1})-(\ref{case1estimate3}) we obtain (\ref{PperpJaHaJaPperp_below}). 

\textbf{Obtaining estimate (\ref{case1estimate1}):} Note that the clusters are missing $\lvert{\mathcal{E}_{k+1}}\rvert$ electrons in order to be neutral. Hence, the set of ionic charges $I = (\delta_1, \dots, \delta_k)$ satisfies
\begin{equation}
    \sum_{j=1}^k \lvert{\delta_j}\rvert \neq 0, \qquad \delta_j \leqslant \sum_{l \in \mathcal{N}_j} Z_l, \qquad \sum_{j=1}^k \delta_j = \lvert{\mathcal{E}_{k+1}}\rvert. \label{clusterionsmissingelectrons}
\end{equation}
The last condition represents that $\lvert{\mathcal{E}_{k+1}}\rvert$ electrons are missing from the clusters. Since $\lvert{\mathcal{E}_{k+1}}\rvert \geqslant 1$, there exists a subgroup of clusters of strictly positive charge, which without loss of generality we index $1 \leqslant j \leqslant n$ and $n < k$, such that 
\begin{equation}
    \sum_{j=1}^n \delta_j \geqslant \lvert{\mathcal{E}_{k+1}}\rvert, \qquad \delta_j > 0, \:\forall\: 1 \leqslant j \leqslant n. 
\end{equation}
Now, consider adding the missing $\lvert{\mathcal{E}_{k+1}}\rvert$ electrons back to these strictly positive ions. These ions are given new charges $\delta_j' \leqslant \delta_j$ with
\begin{equation}
    \sum_{j=1}^n E_{j,0}^{\delta_j} > \sum_{j=1}^n E_{j,0}^{\delta_j'}, \qquad \sum_{j=1}^n \delta_j' = \sum_{j=1}^n \delta_j - \lvert{\mathcal{E}_{k+1}}\rvert. 
\end{equation}
We denote the new set of ionic charges by $I = (\delta_1', \dots, \delta_n', \delta_{n+1}, \dots, \delta_k)$, and in particular in $I'$ we have all the electrons localized near clusters (note these charges could all be zero). Then we have 
\begin{equation} \label{clusterionsmissingelectrons_conclusion}
    E_{\infty, 0}^I \geqslant E_{\infty, 0}^{I'} + G_3 \geqslant E_{\infty, 1} + G.
\end{equation}
So estimating $H_\mathcal{E} \geqslant E_{\infty, 0}^I$ and using (\ref{clusterionsmissingelectrons_conclusion}) we obtain (\ref{case1estimate1}). 

\textbf{Obtaining estimate (\ref{case1estimate2}):} From (\ref{clusterinteraction}) we have
\begin{align}
    &J_{b,n} I_{\mathcal{E}} J_{b,n} \geqslant - \left(\tau_n^b\right)^{-1} \left( \sum_{l \in \mathcal{E}_i} \sum_{s \in \mathcal{N}_j} \frac{e^2 Z_s}{\lvert{\tilde{x}_l - Y_s^n + r_{ij}^n}\rvert} + \sum_{l \in \mathcal{N}_i} \sum_{s \in \mathcal{E}_j} \frac{e^2 Z_l}{\lvert{Y^n_l - \tilde{x}_s + r_{ij}^n}\rvert} \right) \tau_n^b J_{b,n}^2.
\end{align}
By the localization (\ref{R=R(n)choice}) we have $\lvert{\tilde{x}_l - Y^n_s}\rvert \leqslant \frac{r_{ij}^n}{2}$ and so 
\begin{equation}
    \lvert{\tilde{x}_l - Y_s^n + r_{ij}^n}\rvert \geqslant r_{ij}^n - \lvert{\tilde{x}_l - Y^n_s}\rvert \geqslant \frac{r_{ij}^n}{2}.
\end{equation}
Arguing in this way we obtain
\begin{equation}
    \frac{1}{\lvert{\tilde{x}_l - Y_s^n + r_{ij}^n}\rvert} \leqslant \frac{2}{r_{ij}^n}, \qquad \frac{1}{\lvert{Y^n_l - \tilde{x}_s + r_{ij}^n}\rvert} \leqslant \frac{2}{r_{ij}^n},
\end{equation}
on $\supp \tau_n^b J_{b,n}$ and so we obtain (\ref{case1estimate2}) via
\begin{equation} 
    J_{b,n} I_\mathcal{E} J_{b,n} \geqslant - o(1) J_{b,n}^2.
\end{equation}

\textbf{Obtaining estimate (\ref{case1estimate3}):} Observe that 
\begin{equation}
    \sum_{l \in \mathcal{E}_{k+1}} \left(-\frac{1}{2m_e} \Delta_{x_l} + I_l(Y^n) \right) \geqslant - \left(\tau_n^b\right)^{-1} \left( \sum_{j=1}^{k+1} \sum_{\substack{s \in \mathcal{N}_j \\ l \in \mathcal{E}_{k+1}}} \frac{e^2 Z_s}{\lvert{Y^n_s - x_l + z^n_j}\rvert} \right) \tau_n^b.
\end{equation}
On $\supp \tau_n^b J_{b,n}$, by the localization (\ref{R=R(n)choice}) we again have
\begin{equation}
    \lvert{x_l - (Y^n_s + z^n_j)}\rvert \geqslant (R - 1) C_j > 0,
\end{equation}
and hence
\begin{equation}
    \frac{1}{\lvert{Y^n_s - x_l + z^n_j}\rvert} \leqslant \frac{1}{(R - 1) C_j},
\end{equation}
from which it follows that
\begin{equation}
    \sum_{l \in \mathcal{E}_{k+1}} P^\perp J_{b,n} (-\frac{1}{2m_e} \Delta_{x_l} + I_l(Y^n)) J_{b,n} P^\perp \geqslant - o(1) P^\perp J_{b,n}^2 P^\perp. \label{Ek+1_case1}
\end{equation}
From (\ref{Ek+1_case1}) and  $\sum_{\substack{i < j \\ i,j \in \mathcal{E}_{k+1}}} \frac{e^2}{\lvert{x_i - x_j}\rvert} \geqslant 0$ we obtain (\ref{case1estimate3}). \\ 

\subsubsection{Case II} $\lvert{\mathcal{E}_{k+1}}\rvert = 0$. All electrons are localized around a cluster. Then, using the change of variables \eqref{Hbo=Hcluster+Interaction} and using estimate (\ref{case1estimate2}) we write
\begin{equation} 
    P^\perp J_{b,n} \tilde{H}_{N,M}(y^n) J_{b,n} P^\perp \geqslant P^\perp J_{b,n} H_{\mathcal{E}}(Y^n) J_{b,n} P^\perp - o(1) P^\perp J_{b,n}^2 P^\perp.
\end{equation}
Hence, we seek to show 
\begin{equation} \label{Ha_case2}
    P^\perp J_{b,n} H_{\mathcal{E}}(Y^n) J_{b,n} P^\perp \geqslant (E_{\infty, 1}(Y^n) + G - o(1)) P^\perp J_{b,n}^2 P^\perp - o(1). 
\end{equation} 

\textbf{Case II.1.a}. The partition $\mathcal{E} = \{\mathcal{E}_1, \dots, \mathcal{E}_k\}$ contains ions and $I=(\delta_1, \dots, \delta_k)$ given by (\ref{Ioncharges_def2}) is not an ionic minimizer. Then, the gap is immediate as $E_{\infty,0}^I = E_{\infty, 1} + G_4$, so
\begin{equation}
    H_{\mathcal{E}}(Y^n) \geqslant E_{\infty,0}^I(Y^n) = E_{\infty, 1}(Y^n) + G
\end{equation}
and we obtain (\ref{Ha_case2}). \\

\textbf{Case II.1.b}. The partition $\mathcal{E} = \{\mathcal{E}_1, \dots, \mathcal{E}_k\}$ contains ions and $I=(\delta_1, \dots, \delta_k)$ given by (\ref{Ioncharges_def2}) is an ionic minimizer. In this case, the arguments used previously will not yield a gap, so instead we proceed as follows.

Denote by $S(\mathcal{E})$ the set of permutations $\pi$ that keeps the partition $\mathcal{E}$ invariant. Let $S_\mathcal{E}$ denote the corresponding symmetrizer, i.e.
\begin{equation}
    S_\mathcal{E} = \frac{1}{\lvert{S(\mathcal{E})}\rvert} \sum_{\pi \in S(\mathcal{E})} T_\pi,
\end{equation}
where $\lvert{S(\mathcal{E})}\rvert = \prod_{j=1}^k \lvert{\mathcal{E}_j}\rvert !$. We remark that $S_\mathcal{E} \psi^{\mathcal{E},n}_I = \psi^{\mathcal{E},n}_I$ and so $\lVert{S_\mathcal{E} \psi^{\mathcal{E},n}_I}\rVert = \lVert{\psi^{\mathcal{E},n}_I}\rVert = 1$. Then, define the orthogonal projection
\begin{equation}
    P_I \phi = \psi^{\mathcal{E},n}_I \langle \psi^{\mathcal{E},n}_I , \phi \rangle,
\end{equation}
with the orthogonal complement $P_I^\perp = 1 - P_I$. By construction, $P_I$ and $P_I^\perp$ are spectral projections of $H_{\mathcal{E}}$. Introducing $1 = P_I^\perp + P_I$, we write
\begin{align}
    P^\perp J_{b,n} H_{\mathcal{E}} J_{b,n} P^\perp &=  P^\perp J_{b,n} P_I^\perp H_{\mathcal{E}} P_I^\perp J_{b,n} P^\perp + P^\perp J_{b,n} P_I H_{\mathcal{E}} P_I J_{b,n} P^\perp, \label{Ha_case3_1}
\end{align}
and we prove below the estimates
\begin{align}
    P^\perp J_{b,n} P_I^\perp H_{\mathcal{E}} P_I^\perp J_{b,n} P^\perp &\geqslant (E_{\infty,1} + G) P^\perp J_{b,n}^2 P^\perp, \label{Ha_case3_estimate1} \\
    P^\perp J_{b,n} P_I H_{\mathcal{E}} P_I J_{b,n} P^\perp &\geqslant -o(1) P^\perp J_{b,n}^2 P^\perp. \label{Ha_case3_estimate2}
\end{align}
Combining (\ref{Ha_case3_1}), (\ref{Ha_case3_estimate1}), and (\ref{Ha_case3_estimate2}) we obtain (\ref{Ha_case2}). 

\par
\textbf{Obtaining the estimate (\ref{Ha_case3_estimate1}):} First, define the orthogonal projections 
\begin{equation}
    P_j \phi = S_{j,b} \psi_{\mathcal{E}_j,0}^{\delta_j, (n)} \langle S_{j,b} \psi_{\mathcal{E}_j,0}^{\delta_j, (n)}, \phi \rangle_{x_{\mathcal{E}_j}},
\end{equation}
where $S_{j,b}$ is the symmetrizer over all permutations that act only on the coordinates $\mathcal{E}_j$ (and hence $S_{j,b} \psi_{\mathcal{E}_j,0}^{\delta_j, (n)} = \psi_{\mathcal{E}_j,0}^{\delta_j, (n)}$ and $S_{j,b} J_{b,n} = J_{b,n}$). Then we have
\begin{equation}
    P_I = \prod_{j=1}^k P_j,
\end{equation}
since $S_1 \dots S_k = S_\mathcal{E}$ and $\bigotimes_{j=1}^k \psi_{\mathcal{E}_j,0}^{\delta_j, (n)} = \psi^{\mathcal{E},n}_I$. It can be proven by induction on $k$ that 
\begin{equation} \label{PIperp_induction}
    P_I^\perp = 1 - \prod_{j=1}^k P_{j} = \sum_{i=1}^k \left( P_i^\perp \prod_{j = i+1}^k P_{j} \right).
\end{equation}
Then, since $P_{j}^\perp P_{j} = 0 = P_{j} P_{j}^\perp$ and $[P_{j}, H_{\mathcal{E}}] = 0$, writing $H_{\mathcal{E}} = \sum_{j=1}^k H_{\mathcal{E}_j}$ we have
\begin{align}
    P_I^\perp H_{\mathcal{E}} P_I^\perp &= (\sum_{i=1}^k P_i^\perp \prod_{j = i+1}^k P_{j}) H_{\mathcal{E}} (\sum_{l=1}^k P_l^\perp \prod_{r = l+1}^k P_{r}) \nonumber \\
    &= \sum_{i=1}^k \left( \prod_{l = i+1}^k P_l \right) P_i^\perp H_{\mathcal{E}} P_i^\perp \left( \prod_{l = i+1}^k P_l \right) \nonumber \\
    &= \sum_{j=1}^k \sum_{i=1}^k \left( \prod_{l = i+1}^k P_l \right) P_i^\perp H_{\mathcal{E}_j} P_i^\perp \left( \prod_{l = i+1}^k P_l \right). \label{Paperp_Ha_Paperp_0}
\end{align}
For each $j$, if $i=j$ we estimate $P_i^\perp H_{\mathcal{E}_j} P_i^\perp \geqslant E_{j, 0}^{\delta_j} + G_2$ (as $P_j^\perp$ removes the ground state energy from the spectrum of $H_{\mathcal{E}_j}$), and if $i \neq j$ we simply estimate $H_{\mathcal{E}_j} \geqslant E_{j, 0}^{\delta_j}$. Thus, we obtain from (\ref{Paperp_Ha_Paperp_0}) that 
\begin{align}
     P_I^\perp H_{\mathcal{E}} P_I^\perp \geqslant \sum_{i=1}^k \left(\sum_{j=1}^k E_{j,0}^{\delta_j} + G_2 \right)  P_i^\perp \left( \prod_{l = i+1}^k P_l \right) &\geqslant (E_{\infty,0}^I + G_2) P_I^\perp  \nonumber \\
     &\geqslant (E_{\infty, 1} + G) P_I^\perp. \label{Paperp_Ha_Paperp_1}
\end{align}
Returning to the left-hand side of (\ref{Ha_case3_estimate1}), we have
\begin{align}
    \langle \phi, P^\perp J_{b,n} P_I^\perp H_{\mathcal{E}} P_I^\perp J_{b,n} P^\perp \phi \rangle &\geqslant (E_{\infty,1} + G) \lVert{P_I^\perp J_{b,n} P^\perp \phi}\rVert^2 \geqslant (E_{\infty,1} + G) \lVert{J_{b,n} P^\perp \phi}\rVert^2,
\end{align}
since we have already established (\ref{E0+C<0}) and so we obtain (\ref{Ha_case3_estimate1}).

\textbf{Obtaining the estimate (\ref{Ha_case3_estimate2}):} Using $H_{\mathcal{E}} P_I = E_{\infty,0}^I P_I$ we write
\begin{align}
    \langle \phi, P^\perp J_{b,n} P_I H_{\mathcal{E}} P_I J_{b,n} P^\perp \phi \rangle = E_{\infty,0}^I \lVert{P_I J_{b,n} P^\perp \phi}\rVert^2 &= E_{\infty,0}^I \lvert{\langle \psi^{\mathcal{E},n}_I, J_{b,n} P^\perp \phi \rangle}\rvert^2 \nonumber \\
    &= E_{\infty,0}^I \lvert{\langle J_{b,n} \psi^{\mathcal{E},n}_I, P^\perp \phi \rangle}\rvert^2. \label{Ha_case3_1'}
\end{align}
First we show that $(1-J_{b,n})\psi^{\mathcal{E},n}_I = o(1)$, so that
\begin{equation}
    \langle J_{b,n} \psi^{\mathcal{E},n}_I, P^\perp \phi \rangle = \langle \psi^{\mathcal{E},n}_I, P^\perp \phi \rangle + o(1). \label{Ha_case3_2}
\end{equation} 
Introducing $1 = e^{-\alpha \sum_{j=1}^k \langle \tilde{x}_{\mathcal{E}_j} \rangle} e^{\alpha \sum_{j=1}^k \langle \tilde{x}_{\mathcal{E}_j} \rangle}$ and using (\ref{exponentiallocalizationofeigenfunctions}), we have
\begin{equation}
    \lVert{(1-J_{b,n}) \psi^{\mathcal{E},n}_I}\rVert \lesssim \lVert{(1-J_{b,n}) e^{-\alpha \sum_{j=1}^k \langle \tilde{x}_{\mathcal{E}_j} \rangle}}\rVert_{\infty}.
\end{equation}
Since $1 \leqslant 1 + J_{b,n}$, we have
\begin{align}
    (1-J_{b,n}) e^{-\alpha \sum_{j=1}^k \langle \tilde{x}_{\mathcal{E}_j} \rangle} &\leqslant \left(1 - J_{b,n}^2 \right) e^{-\alpha \sum_{j=1}^k \langle \tilde{x}_{\mathcal{E}_j} \rangle} = \sum_{b' \neq b} J_{b', R}^2 e^{-\alpha \sum_{j=1}^k \langle \tilde{x}_{\mathcal{E}_j} \rangle}. 
\end{align}
There is at least one electron such that in the coordinates in the exponent it is localized at the cluster $j$ for some $j$ but on $\supp J_{b',R}$ for $b' \neq b$ it is localized at a different cluster. Thus, according to (\ref{R=R(n)choice}), on $\supp J_{b',R}$ we have
\begin{equation}
     e^{-\alpha \sum_{j=1}^k \langle \tilde{x}_{\mathcal{E}_j} \rangle} \leqslant  e^{-\alpha (1 + 4 R^2 C_j)^{1/2}} = o(1),
\end{equation}
and so 
\begin{equation}
    \lVert{(1-J_{b,n}) e^{-\alpha \sum_{j=1}^k \langle \tilde{x}_{\mathcal{E}_j} \rangle}}\rVert_{\infty} \leqslant \lVert{\sum_{b' \neq b} J_{b', R}^2 e^{-\alpha \sum_{j=1}^k \langle \tilde{x}_{\mathcal{E}_j} \rangle}}\rVert_\infty = o(1),
\end{equation}
which finally implies
\begin{equation}
    \lVert{(1-J_{b,n}) \psi^{\mathcal{E},n}_I}\rVert = o(1). \label{JaR_exponentiallocalization}
\end{equation}
Returning to the right-hand side of \eqref{Ha_case3_2}, we use the fact that $\phi \in \cH_s$, i.e. $S \phi = \phi$, and $[P^\perp, S] = 0$ to see
\begin{equation}
    \langle \psi^{\mathcal{E},n}_I, P^\perp \phi \rangle = \langle S \psi^{\mathcal{E},n}_I, P^\perp \phi \rangle = 0. \label{Ha_case3_2_0}
\end{equation}
Combining \eqref{Ha_case3_2_0}, \eqref{Ha_case3_2}, and \eqref{Ha_case3_1'} yields the desired estimate (\ref{Ha_case3_estimate2}). \\

\textbf{Case II.2}. The partition $\mathcal{E} = \{\mathcal{E}_1, \dots, \mathcal{E}_k\}$ features only neutral clusters. Then we follow the same steps as in the previous case, with the necessary modifications. We define the orthogonal projection
\begin{equation}
    P_\mathcal{E} \phi = \tilde{S}_\mathcal{E}(\psi^{\mathcal{E},n}_{0}) \langle \tilde{S}_\mathcal{E}(\psi^{\mathcal{E},n}_{0}), \phi \rangle + \sum_{l=1}^k \sum_{p = 1}^{m_l} \tilde{S}_\mathcal{E}(\psi^{\mathcal{E},n}_{l,p}) \langle \tilde{S}_\mathcal{E}(\psi^{\mathcal{E},n}_{l,p}), \phi \rangle, 
\end{equation}
where recall $m_l$ denotes the multiplicity of the first excited state. Define the orthogonal complement $P_\mathcal{E}^\perp = 1 - P_\mathcal{E}$. By construction $P_\mathcal{E}$ and $P_\mathcal{E}^\perp$ are spectral projections of $H_\mathcal{E}$. Introducing $P_\mathcal{E} + P_\mathcal{E}^\perp$, we write
\begin{equation} \label{Hacase_II.2}
    P^\perp J_{b,n} H_\mathcal{E} J_{b,n} P^\perp = P^\perp J_{b,n} P_\mathcal{E}^\perp H_\mathcal{E} P_\mathcal{E}^\perp J_{b,n} P^\perp + P^\perp J_{b,n} P_\mathcal{E} H_\mathcal{E} P_\mathcal{E} J_{b,n} P^\perp,
\end{equation}
and we have the estimates
\begin{align}
    P^\perp J_{b,n} P_\mathcal{E}^\perp H_\mathcal{E} P_\mathcal{E}^\perp J_{b,n} P^\perp &\geqslant (E_{\infty, 1} + G) P^\perp J_{b,n}^2 P^\perp, \label{caseII.2_estimate1} \\
    P^\perp J_{b,n} P_\mathcal{E} H_\mathcal{E} P_\mathcal{E} J_{b,n} P^\perp &\geqslant - o(1). \label{caseII.2_estimate2}
\end{align}
Combining (\ref{Hacase_II.2}), (\ref{caseII.2_estimate1}) and (\ref{caseII.2_estimate2}) we obtain (\ref{Ha_case2}). 

\textbf{Obtaining estimate (\ref{caseII.2_estimate1}):} First, define the orthogonal projections
\begin{equation}
    P_{j,0} \phi = \psi_{\mathcal{E}_j,0}^{(n)} \langle \psi_{\mathcal{E}_j,0}^{(n)}, \phi \rangle, \qquad P_{l,1} \phi = \sum_{p = 1}^{m_l} \psi_{\mathcal{E}_l,1,p}^{(n)} \langle \psi_{\mathcal{E}_l,1,p}^{(n)}, \phi \rangle.
\end{equation}
As before, note that $S_\mathcal{E} = S_{1,b} \dots S_{k,b}$ and $\lVert{S_\mathcal{E} \phi }\rVert = \prod_{j=1}^k \lVert{S_{j,b} \phi_j}\rVert$ if $\phi = \otimes_{j=1}^k \phi_j$. So, we write
\begin{equation} \label{Paexpansion}
    P_\mathcal{E} = \prod_{j=1}^k P_{j,0} + \sum_{l=1}^k P_{l,1} \prod_{\substack{j=1 \\ j\neq 1}}^k P_{j,0},
\end{equation}
and using (\ref{PIperp_induction}) we write
\begin{align}
    P_\mathcal{E}^\perp = 1 - \prod_{j=1}^k P_{j,0} - \sum_{l=1}^k P_{l,1} \prod_{\substack{j=1 \\ j\neq 1}}^k P_{j,0} &= \sum_{l=1}^k \left( P_{l,0}^\perp \prod_{j = l+1}^k P_{j,0} - P_{l,1} \prod_{\substack{j=1 \\ j\neq 1}}^k P_{j,0} \right) \nonumber \\
    &= \sum_{l=1}^k \left(P_{l,0}^\perp - P_{l,1} \prod_{j=1}^{l-1} P_{j,0} \right) \prod_{j=l+1}^k P_{j,0}. \label{Paperp_step1}
\end{align}
For $l \geqslant 2$, we write 
\begin{equation}
    P_{l,0}^\perp - P_{l,1} \prod_{j=1}^{l-1} P_{j,0} = \left(P_{l,0}+P_{l,1}\right)^\perp + P_{l,1}\left(\prod_{j=1}^{l-1}P_{j,0}\right)^\perp. \label{Paperp_step2}
\end{equation}
Using (\ref{PIperp_induction}) again, we have
\begin{equation}
    \left(\prod_{j=1}^{l-1}P_{j,0}\right)^\perp = \sum_{j=1}^{l-1} P_{j,0}^\perp \prod_{i=j+1}^{l-1} P_{i,0}. \label{Paperp_step3}
\end{equation}
Combining (\ref{Paperp_step1}), (\ref{Paperp_step2}), (\ref{Paperp_step3}) we obtain 
\begin{align}
    P_\mathcal{E}^\perp &= \sum_{l=1}^k \left[\left(P_{l,0} + P_{l,1} \right)^\perp + P_{l,1} \sum_{j=1}^{l-1} P_{j,0}^\perp \prod_{i = j+1}^{l-1} P_{i,0} \right] \prod_{i=l+1}^k P_{i,0} \nonumber \\
    &= \sum_{l=1}^k \left( P_{l,0} + P_{l,1} \right)^\perp \prod_{j = l+1}^k P_{j,0} + \sum_{l=2}^k \sum_{j=1}^{l-1} P_{l,1} P_{j,0}^\perp \prod_{\substack{i = j+1 \\ i \neq l}}^k P_{i,0}. \label{Paperp}
\end{align}
Using (\ref{Paperp}) and writing $H_\mathcal{E} = \sum_{n=1}^k H_{\mathcal{E}_n}$, analogous to (\ref{Paperp_Ha_Paperp_0}) we obtain
\begin{align}
    P_\mathcal{E}^\perp H_\mathcal{E} P_\mathcal{E}^\perp &= \sum_{n=1}^k \sum_{l=1}^k \left(\prod_{j = l+1}^k P_{j,0}\right) \left( P_{l,0} + P_{l,1} \right)^\perp H_{\mathcal{E}_n} \left( P_{l,0} + P_{l,1} \right)^\perp \left(\prod_{j = l+1}^k P_{j,0}\right) \nonumber \\
    &\hspace{10pt} + \sum_{n=1}^k \sum_{l=2}^k \sum_{j=1}^{l-1} \left(\prod_{\substack{i = j+1 \\ i \neq l}}^k P_{i,0}\right) P_{l,1} P_{j,0}^\perp H_{\mathcal{E}_n} P_{j,0}^\perp P_{l,1} \left(\prod_{\substack{i = j+1 \\ i \neq l}}^k P_{i,0}\right). \label{Paperp_Ha_Paperp__CaseII.2}
\end{align}
Consider the first term. For each $n$, if $l=n$ we estimate 
\begin{equation}
    \left( P_{l,0} + P_{l,1} \right)^\perp H_{\mathcal{E}_n} \left( P_{l,0} + P_{l,1} \right)^\perp \geqslant (E_{j,1} + G_1) \left( P_{l,0} + P_{l,1} \right)^\perp,
\end{equation}
as $\left( P_{l,0} + P_{l,1} \right)^\perp$ removes the ground state energy and the first excited state energy from the spectrum of $H_{\mathcal{E}_n}$, and if $l \neq j$ we simply estimate $H_{\mathcal{E}_n} \geqslant E_{n,0}$. For the second term, if $j=n$ we estimate
\begin{equation}
    P_{i,0}^\perp H_{\mathcal{E}_n} P_{i,0}^\perp \geqslant (E_{n,0} + G_1) P_{i,0}^\perp,
\end{equation}
if $l=n$ we estimate $P_{l,1} H_{\mathcal{E}_n} P_{l,1} = E_{n,1} P_{l,1}$, and else we estimate $H_{\mathcal{E}_n} \geqslant E_{n,0}$. Hence, analogous to (\ref{Paperp_Ha_Paperp_1}) from (\ref{Paperp_Ha_Paperp__CaseII.2}) we obtain (\ref{caseII.2_estimate1}) via
\begin{equation}
    P_\mathcal{E}^\perp H_\mathcal{E} P_\mathcal{E}^\perp \geqslant (E_{\infty, 1} + G) P_\mathcal{E}^\perp. 
\end{equation}

\textbf{Obtaining estimate (\ref{caseII.2_estimate2}):} Denote
\begin{equation}
    P_{\mathcal{E},0} := \prod_{j=1}^k P_{j,0}, \qquad P_{\mathcal{E},l} := P_{l,1} \prod_{\substack{j=1 \\ j\neq 1}}^k P_{j,0}, \qquad 1 \leqslant l \leqslant k,
\end{equation}
so that following (\ref{Paexpansion}) we have $P_\mathcal{E} = \sum_{l=0}^k P_{\mathcal{E}, l}$ and hence
\begin{equation} \label{PaHaPa_step1}
    P^\perp J_{b,n} P_\mathcal{E} H_\mathcal{E} P_\mathcal{E} J_{b,n} P^\perp = \sum_{l=0}^k P^\perp J_{b,n} P_{\mathcal{E},l} H_\mathcal{E} P_{\mathcal{E},l} J_{b,n} P^\perp.
\end{equation}
Then, observe that each $P_{\mathcal{E},l}$ is a spectral projection of $H_\mathcal{E}$, and we have $H_\mathcal{E} P_{\mathcal{E},l} = E_{\infty,0}^1 P_{\mathcal{E},l}$. Hence for each $0 \leqslant l \leqslant k$ the argument to obtain the estimate (\ref{Ha_case3_estimate2}) follows analogously, i.e. we have
\begin{equation}
    P^\perp J_{b,n} P_{\mathcal{E},l} H_\mathcal{E} P_{\mathcal{E},l} J_{b,n} P^\perp \geqslant -o(1).
\end{equation}
Combining this with (\ref{PaHaPa_step1}), we obtain (\ref{caseII.2_estimate2}). 

% This concludes the proof of Proposition \ref{PperpJaHaJaPperp_belowestimates}.
\qed

\subsection{Proof of Lemma \ref{FSMap_estimates}} The main idea in all cases is to use the cluster decomposition and the fact that different eigenfunctions are orthogonal if belonging to the same cluster or otherwise exponentially localized around different clusters. 

\textbf{Proof of (\ref{almostorthogonal_estimate}):} Without loss of generality we will only consider $\phi_i = \tilde{S}(\psi^{\mathcal{E}, n}_0)$ and $\phi_j = \tilde{S}(\psi^{\mathcal{E}, n}_{1,1})$, and we show the estimate
\begin{equation}
    \langle \tilde{S}(\psi^{\mathcal{E}, n}_0), \tilde{S}(\psi^{\mathcal{E}, n}_{1,1}) \rangle = o(1),
\end{equation}
 where $\mathcal{E}$ is some partition of the electrons into neutral clusters. All other cases of \eqref{almostorthogonal_estimate} follow either by normalization, orthogonality, or using the argument presented below. We change variables inside the inner product and expand the symmetrizer (recall (\ref{def_symmetrizer})) to obtain
\begin{align} \label{S0S1_expanded}
    &\langle \tilde{S}(\psi^{\mathcal{E}, n}_0), \tilde{S}(\psi^{\mathcal{E}, n}_{1,1}) \rangle = \frac{1}{(N!)^2 \lVert{S(\psi^{\mathcal{E},n}_0)}\rVert \lVert{S(\psi^{\mathcal{E}, n}_{1,1})}\rVert} \sum_{\pi, \pi'} \langle \psi^{\mathcal{E}(\pi), n}_0, \psi^{\mathcal{E}(\pi'), n}_{1,1} \rangle. 
\end{align}
We write $\mathcal{E} = \mathcal{E}(\pi)$ and $\mathcal{E}' = \mathcal{E}(\pi')$ in order to simplify notation. If $\mathcal{E} = \mathcal{E}'$, then $\psi_{0}^{\mathcal{E},n} \perp \psi_{1,1}^{\mathcal{E},n}$ implies
\begin{equation} \label{psipi_psipi'=0}
    \langle \psi^{\mathcal{E},n}_0, \psi^{\mathcal{E}',n}_{1,1} \rangle = 0.
\end{equation}
If $\mathcal{E} \neq \mathcal{E}'$, then using the exponential localization of eigenfunctions (\ref{exponentiallocalizationofeigenfunctions}), we have
\begin{align}
    \lvert{\langle \psi^{\mathcal{E},n}_0, \psi^{\mathcal{E}',n}_{1,1} \rangle}\rvert &= \lvert{\langle e^{\alpha \sum \langle \tilde{x}_{\mathcal{E}_j} \rangle} \psi^{\mathcal{E},n}_0, e^{- \alpha (\sum \langle \tilde{x}_{\mathcal{E}_j} \rangle + \sum \langle \tilde{x}_{\mathcal{E}_j'}' \rangle)} e^{\alpha \sum \langle \tilde{x}_{\mathcal{E}_j'}' \rangle} \psi^{\mathcal{E}',n}_{1,1} \rangle }\rvert \nonumber \\
    &\lesssim \lVert{e^{- \alpha (\sum \langle \tilde{x}_{\mathcal{E}_j} \rangle + \sum \langle \tilde{x}_{\mathcal{E}_j'}' \rangle)}}\rVert_\infty,
\end{align}
where $\tilde{x}_{\mathcal{E}_j}$ is the tuple of electron coordinates localized at cluster $j$ in the partition $\mathcal{E}$, and likewise $\tilde{x}_{\mathcal{E}_j'}'$ in the partition $\mathcal{E}'$. Since $\mathcal{E} \neq \mathcal{E}'$ there is at least one electron localized in different clusters in the partitions $\mathcal{E}$ and $\mathcal{E}'$. Without loss of generality let this electron be $x_1$ localized at cluster $j$ in partition $\mathcal{E}$ and at cluster $l$ in partition $\mathcal{E}'$. Then, using the triangle inequality for the $\langle \cdot \rangle$ bracket, we have
\begin{align}
    \sum \langle \tilde{x}_{\mathcal{E}_j} \rangle + \sum \langle \tilde{x}_{\mathcal{E}_j'}' \rangle &\geqslant \langle \tilde{x}_{\mathcal{E}_j} \rangle + \langle \tilde{x}_{\mathcal{E}_l'}' \rangle \geqslant \langle \tilde{x}_1 \rangle + \langle \tilde{x}_1' \rangle \geqslant \langle r_{lj}^n \rangle. 
\end{align}
Thus,
\begin{equation}
    \lVert{e^{- \alpha (\sum \langle \tilde{x}_{\mathcal{E}_j} \rangle + \sum \langle \tilde{x}_{\mathcal{E}_j'}' \rangle)}}\rVert_\infty \leqslant \lVert{e^{- \alpha \langle r_{lj}^n \rangle }}\rVert_\infty = o(1),
\end{equation}
and hence
\begin{equation} \label{psipi_psipi'=o(1)}
    \langle \psi^{\mathcal{E},n}_0, \psi^{\mathcal{E}',n}_{1,1} \rangle = o(1). 
\end{equation}
Combining (\ref{S0S1_expanded}), (\ref{psipi_psipi'=0}), and (\ref{psipi_psipi'=o(1)}) we obtain (\ref{almostorthogonal_estimate}).

\textbf{Proof of (\ref{SHboS_estimate}):} As above, let us consider the example case $\phi_i = \phi_j = \tilde{S}(\psi^{\mathcal{E},n}_0)$ where $\mathcal{E}$ is some partition of the electrons into neutral clusters. The other cases will follow analogously. Using (\ref{def_symmetrizer}) and (\ref{Hbo=Hcluster+Interaction}) - \eqref{clustertranslation}, we write
\begin{align}
     &\langle \tilde{S}(\psi^{\mathcal{E},n}_0), \tilde{H}_{N,M}(y^n) \tilde{S}(\psi^{\mathcal{E},n}_0) \rangle \nonumber \\
     &= \frac{1}{N! \lVert{S(\psi^{\mathcal{E},n}_0)}\rVert} \sum_{\pi} \langle \tilde{S}(\psi^{\mathcal{E},n}_0), \tilde{H}_{N,M}(y^n) T_\pi \psi^{\mathcal{E},n}_0 \rangle \nonumber \\
     &= \frac{1}{N! \lVert{S(\psi^{\mathcal{E},n}_0)}\rVert} \sum_{\pi} \langle \tilde{S}(\psi^{\mathcal{E},n}_0), H_{\mathcal{E}(\pi)}(Y^n) \psi^{\mathcal{E}(\pi), n}_0 \rangle \nonumber \\
     &\hspace{20pt} + \frac{1}{N! \lVert{S(\psi^{\mathcal{E},n}_0)}\rVert} \sum_{\pi} \sum_{i < l} \langle \tilde{S}(\psi^{\mathcal{E},n}_0), I_{\mathcal{E}(\pi)_i, \mathcal{E}(\pi)_l}(Y_{\mathcal{N}_i}^n, Y_{\mathcal{N}_l}^n) \psi^{\mathcal{E}(\pi), n}_0 \rangle \nonumber \\
     &= E_{\infty,0} + \frac{1}{N! \lVert{S(\psi^{\mathcal{E},n}_0)}\rVert} \sum_{\pi} \sum_{i < l} \langle \tilde{S}(\psi^{\mathcal{E},n}_0), I_{\mathcal{E}(\pi)_i, \mathcal{E}(\pi)_l}(Y_{\mathcal{N}_i}^n, Y_{\mathcal{N}_l}^n)  \psi^{\mathcal{E}(\pi), n}_0 \rangle. \label{E_0(yn)expansion2}
\end{align}
It remains to estimate the second term on the RHS of (\ref{E_0(yn)expansion2}). By the CS inequality it is enough to estimate $\lVert{I_{\mathcal{E}(\pi)_i, \mathcal{E}(\pi)_l}(Y_{\mathcal{N}_i}^n, Y_{\mathcal{N}_l}^n) \psi^{\mathcal{E}(\pi), n}_0}\rVert$. To simplify notation, we drop the $\pi$ in $\mathcal{E}(\pi)$ and write simply $\mathcal{E}$. Using (\ref{clusterinteraction}) we expand to obtain
\begin{align}
    I_{\mathcal{E}_i, \mathcal{E}_l}(Y_{\mathcal{N}_i}^n, Y_{\mathcal{N}_l}^n) \tau_n \psi^{\mathcal{E}, n}_0(x) &= \tau_n^{-1} \frac{1}{2} \sum_{m \in \mathcal{E}_i} \sum_{s \in \mathcal{E}_l} \frac{e^2}{\lvert{\tilde{x}_m - \tilde{x}_s + r_{il}^n}\rvert} \tau_n \psi^{\mathcal{E}, n}_0(x) \nonumber \\
    &\hspace{10pt} - \tau_n^{-1} \sum_{m \in \mathcal{E}_i} \sum_{s \in \mathcal{N}_l} \frac{e^2 Z_s}{\lvert{\tilde{x}_m - Y_s^n + r_{il}^n}\rvert} \tau_n \psi^{\mathcal{E}, n}_0(x) \nonumber \\
    &\hspace{10pt} - \tau_n^{-1} \sum_{m \in \mathcal{N}_i} \sum_{s \in \mathcal{E}_l} \frac{e^2 Z_m}{\lvert{Y^n_m - \tilde{x}_s + r_{il}^n}\rvert} \tau_n \psi^{\mathcal{E}, n}_0(x), \label{E_0(yn)expansion3}
\end{align}
and estimate 
\begin{align}
    \lVert{\frac{1}{\lvert{\tilde{x}_m - \tilde{x}_s + r_{ij}^n}\rvert} \bigotimes_{j=1}^k \psi_{\mathcal{E}_j, 0}^{(n)}(\tilde x_{\mathcal{E}_j})}\rVert^2 &\leqslant o(1), \label{interactionterms_ee_decay}\\
    \lVert{\frac{1}{\lvert{\tilde{x}_m - Y_s^n + r_{ij}^n}\rvert} \bigotimes_{j=1}^k \psi_{\mathcal{E}_j, 0}^{(n)}(\tilde x_{\mathcal{E}_j})}\rVert^2 &\leqslant o(1), \label{interactionterms_en_decay}
\end{align}
where for $m \in \mathcal{E}_i$ and $i\neq l$, we write $\tilde{x}_s$ for $s \in \mathcal{E}_l$ and $Y^n_s$ for $s \in \mathcal{N}_l$. 

The estimates (\ref{interactionterms_ee_decay}) and (\ref{interactionterms_en_decay}) together with the expansion (\ref{E_0(yn)expansion3}) imply that 
\begin{equation}
    \lVert{I_{\mathcal{E}(\pi)_i, \mathcal{E}(\pi)_l}^{\sigma(\pi)} \psi^{\mathcal{E}(\pi), n}_0}\rVert = o(1).
\end{equation}
From (\ref{E_0(yn)expansion2}) we obtain (\ref{SHboS_estimate}).

If instead we considered $\phi_i \neq \phi_j$, then we would proceed as follows. Consider the example case $\phi_i = \tilde{S}(\psi^{\mathcal{E},n}_0)$ and $\phi_j = \tilde{S}(\psi^{\mathcal{E},n}_{1,1})$, where $\mathcal{E}$ is a partition of the electrons into neutral clusters. Then as before, using (\ref{def_symmetrizer}), (\ref{Hbo=Hcluster+Interaction}), (\ref{Hboclusters}), as well as the estimates (\ref{E_0(yn)expansion3})-(\ref{interactionterms_en_decay}) from which it follows that $\lVert{I_{\mathcal{E}(\pi)}(Y^n) \tilde{S}(\psi^{\mathcal{E}(\pi), n}_{1,1})}\rVert = o(1)$, we write
\begin{equation} \label{S0HboS1_expanded}
    \langle \tilde{S}(\psi^{\mathcal{E},n}_0), \tilde{H}_{N,M}(y^n) \tilde{S}(\psi^\mathcal{E}_1) \rangle = E_{\infty,0}^1 \langle \tilde{S}(\psi^{\mathcal{E},n}_0), \tilde{S}(\psi^{\mathcal{E},n}_{1,1}) \rangle + o(1),
\end{equation}
and \eqref{SHboS_estimate} follows from \eqref{almostorthogonal_estimate}.

\textbf{Obtaining estimate (\ref{interactionterms_ee_decay}):} Recall that the $\sim$ on the $\tilde x_{\mathcal{N}_i}$ coordinates means that they are centered at the center $z^n_i$ of the cluster $\mathcal{N}_i$, i.e. $\phi(\tilde x) = \tau_n \phi(x)$. By the normalization of the ground states,
\begin{equation} \label{ee_interaction_bound1}
    \lVert{\frac{1}{\lvert{\tilde x_m - \tilde x_s + r_{il}^n}\rvert} \bigotimes_{j=1}^k \psi_{\mathcal{E}_j, 0}^{(n)}(\tilde x_{\mathcal{E}_j})}\rVert^2 = \int \frac{\lvert{\psi_{\mathcal{E}_i,0}^{(n)}(\tilde x_{\mathcal{E}_i}) \psi_{\mathcal{E}_l,0}^{(n)}(\tilde x_{\mathcal{E}_l})}\rvert^2}{\lvert{\tilde x_m - \tilde x_s + r_{il}^n}\rvert^2} d \tilde x_{\mathcal{E}_i} d \tilde x_{\mathcal{E}_l}, 
\end{equation}
where recall $m \in \mathcal{E}_i$ and $s \in \mathcal{E}_l$, $i\neq l$. For $\lvert{\tilde x_m}\rvert \leqslant \frac{r_{ij}^n}{4}$ and $\lvert{\tilde x_s}\rvert \leqslant \frac{r_{il}^n}{4}$, we have
\begin{equation}
    \lvert{\tilde x_m - \tilde x_s + r_{il}^n}\rvert \geqslant r_{il}^n - \lvert{\tilde x_m}\rvert - \lvert{\tilde x_s}\rvert \geqslant \frac{r_{il}^n}{2}, 
\end{equation}
and hence
\begin{equation}
    \frac{1}{\lvert{\tilde x_m - \tilde x_s + r_{il}^n}\rvert^2} \leqslant \frac{4}{\left(r_{il}^n\right)^2}.
\end{equation}
Using this and the normalization of the ground states we obtain
\begin{equation} \label{ee_interaction_bound2}
    \int_{\substack{\{\lvert{\tilde x_m}\rvert \leqslant \frac{r_{il}^n}{4}\} \\ \cap \{\lvert{\tilde x_s}\rvert \leqslant \frac{r_{il}^n}{4}\} }} \frac{\lvert{\psi_{\mathcal{E}_i, 0}^{(n)}(\tilde x_{\mathcal{E}_i}) \psi_{\mathcal{E}_l, 0}^{(n)}(\tilde x_{\mathcal{E}_l})}\rvert^2}{\lvert{\tilde x_m - \tilde x_s + r_{il}^n}\rvert^2} d \tilde x_{\mathcal{E}_i} d \tilde x_{\mathcal{E}_l} \leqslant \frac{4}{\left(r_{il}^n \right)^2}. 
\end{equation}
On the domain $(\{\lvert{\tilde x_m}\rvert \leqslant \frac{r_{il}^n}{4}\} \cap \{\lvert{\tilde x_s}\rvert \leqslant \frac{r_{il}^n}{4}\})^c = \{\lvert{\tilde x_m}\rvert > \frac{r_{il}^n}{4}\} \cup \{\lvert{\tilde x_s}\rvert > \frac{r_{il}^n}{4}\}$, we split the integral into two pieces, 
\begin{align}
    \int_{\substack{\{\lvert{\tilde x_m}\rvert > \frac{r_{il}^n}{4}\} \\ \cup \{\lvert{\tilde x_s}\rvert > \frac{r_{il}^n}{4}\}}} \frac{\lvert{\psi_{\mathcal{E}_i, 0}^{(n)}(\tilde x_{\mathcal{E}_i}) \psi_{\mathcal{E}_l, 0}^{(n)}(\tilde x_{\mathcal{E}_l})}\rvert^2}{\lvert{\tilde x_m - \tilde x_s + r_{il}^n}\rvert^2} d \tilde x_{\mathcal{E}_i} d \tilde x_{\mathcal{E}_l} &\leqslant \int_{\{\lvert{\tilde x_m}\rvert > \frac{r_{il}^n}{4}\}} \frac{\lvert{\psi_{\mathcal{E}_i, 0}^{(n)}(\tilde x_{\mathcal{E}_i}) \psi_{\mathcal{E}_l, 0}^{(n)}(\tilde x_{\mathcal{E}_l})}\rvert^2}{\lvert{\tilde x_m - \tilde x_s + r_{il}^n}\rvert^2} d \tilde x_{\mathcal{E}_i} d \tilde x_{\mathcal{E}_l} \nonumber \\
    &\hspace{15pt} + \int_{\{\lvert{\tilde x_s}\rvert > \frac{r_{il}^n}{4}\}} \frac{\lvert{\psi_{\mathcal{E}_i, 0}^{(n)}(\tilde x_{\mathcal{E}_i}) \psi_{\mathcal{E}_l, 0}^{(n)}(\tilde x_{\mathcal{E}_l})}\rvert^2}{\lvert{\tilde x_m - \tilde x_s + r_{il}^n}\rvert^2} d \tilde x_{\mathcal{E}_i} d \tilde x_{\mathcal{E}_l} \label{ee_interaction_bound3}
\end{align}
and, without loss of generality, consider only the first term on the right-hand side. Using the Hardy inequality,
\begin{align}
    \int_{\{\lvert{\tilde x_m}\rvert > \frac{r_{il}^n}{4}\}} \frac{\lvert{\psi_{\mathcal{E}_i, 0}^{(n)}(\tilde x_{\mathcal{E}_i}) \psi_{\mathcal{E}_l, 0}^{(n)}(\tilde x_{\mathcal{E}_l})}\rvert^2}{\lvert{\tilde x_m - \tilde x_s + r_{il}^n}\rvert^2} d \tilde x_{\mathcal{E}_i} d \tilde x_{\mathcal{E}_l} &\leqslant 4 \int_{\{\lvert{\tilde x_m}\rvert > \frac{r_{il}^n}{4}\}} \lvert{\psi_{\mathcal{E}_i, 0}^{(n)}(\tilde x_{\mathcal{E}_i}) \nabla_{\tilde x_s} \psi_{\mathcal{E}_l, 0}^{(n)}(\tilde x_{\mathcal{E}_l})}\rvert^2 d \tilde x_{\mathcal{E}_i} d \tilde x_{\mathcal{E}_l} \nonumber \\
    &\lesssim \int_{\{\lvert{\tilde x_m}\rvert > \frac{r_{il}^n}{4}\}} \lvert{\psi_{\mathcal{E}_i, 0}^{(n)}(\tilde x_{\mathcal{E}_i})}\rvert^2 d \tilde x_{\mathcal{E}_i} \nonumber \\
    &\lesssim e^{-2 \alpha \sqrt{1 + \frac{\left( r_{il}^n\right)^2}{16}}}, \label{ee_interaction_bound4}
\end{align}
where in the last step we used (\ref{exponentiallocalizationofeigenfunctions}) and $e^{-2 \alpha \langle{\tilde x_{\mathcal{E}_i}}\rangle} < e^{-2 \alpha \sqrt{1 + \frac{\left( r_{il}^n \right)^2}{16}}}$ when $\lvert{\tilde x_m}\rvert > \frac{r_{il}^n}{4}$. Combining (\ref{ee_interaction_bound1})-(\ref{ee_interaction_bound4}) we obtain (\ref{interactionterms_ee_decay}). 

\textbf{Obtaining the estimate (\ref{interactionterms_en_decay}):}
Again by normalization,
\begin{equation}
    \lVert{\frac{1}{\lvert{\tilde x_m - Y_s^n + r_{il}^n}\rvert} \bigotimes_{j=1}^k \psi_{\mathcal{E}_j, 0}^{(n)}(\tilde x_{\mathcal{E}_j})}\rVert^2 = \int \frac{\lvert{\psi_{\mathcal{E}_i, 0}^{(n)}(\tilde x_{\mathcal{E}_i})}\rvert^2}{\lvert{\tilde x_m - Y_s^n + r_{il}^n}\rvert^2} d \tilde x_{\mathcal{E}_i}, \label{ee_interaction_bound5}
\end{equation}
where $m \in \mathcal{E}_i$ and $s \in \mathcal{N}_l$, $i \neq l$. For large enough $n$, $\lvert{Y_s^n}\rvert \leqslant \frac{r_{il}^n}{4}$, hence
\begin{equation}
    \int_{\{\lvert{\tilde x_m}\rvert \leqslant \frac{r_{il}^n}{4}\}} \frac{\lvert{\psi_{\mathcal{E}_i, 0}^{(n)}(\tilde x_{\mathcal{E}_i})}\rvert^2}{\lvert{\tilde x_m - Y_s^n + r_{il}^n}\rvert^2} d \tilde x_{\mathcal{E}_i} \leqslant \frac{4}{\left(r_{il}^n \right)^2}. \label{ee_interaction_bound6}
\end{equation}
Using (\ref{exponentiallocalizationofeigenfunctions}), the Hardy inequality, and the boundedness of the derivatives of $e^{\alpha \langle \tilde x_{\mathcal{N}_i}\rangle}$ we estimate the integral on the complement in the following way, 
\begin{align}
    \int_{\{\lvert{\tilde x_m}\rvert > \frac{r_{il}^n}{4}\}} \frac{\lvert{\psi_{\mathcal{E}_i, 0}^{(n)}(\tilde x_{\mathcal{E}_i})}\rvert^2}{\lvert{\tilde x_m - Y_s^n + r_{il}^n}\rvert^2} d \tilde x_{\mathcal{E}_i}
    &\leqslant e^{-2 \alpha \sqrt{1 + \frac{\left(r_{il}^n \right)^2}{16}}} \int_{\{\lvert{\tilde x_m}\rvert > \frac{r_{il}^n}{4}\}} \frac{\lvert{\psi_{\mathcal{E}_i, 0}^{(n)}(\tilde x_{\mathcal{E}_i}) e^{\alpha \langle{\tilde x_{\mathcal{N}_i}}\rangle}}\rvert^2}{\lvert{\tilde x_m - Y_s^n + r_{il}^n}\rvert^2} d \tilde x_{\mathcal{E}_i} \nonumber \\
    &\leqslant 4 e^{-2 \alpha \sqrt{1 + \frac{\left( r_{il}^n\right)^2}{16}}} \int \lvert{\nabla_{\tilde x_m} \psi_{\mathcal{E}_i, 0}^{(n)}(\tilde x_{\mathcal{E}_i}) e^{\alpha \langle{\tilde x_{\mathcal{N}_i}}\rangle}}\rvert^2 d \tilde x_{\mathcal{E}_i} \nonumber \\
    &\lesssim e^{-2 \alpha \sqrt{1 + \frac{\left(r_{il}^n \right)^2}{16}}}. \label{ee_interaction_bound7}
\end{align}
The estimates (\ref{ee_interaction_bound5})-(\ref{ee_interaction_bound7}) imply (\ref{interactionterms_en_decay}).

\textbf{Proof of (\ref{FSresolventterm_estimate}):} Without loss of generality we consider the case $\phi_i = \phi_j = \tilde{S}(\psi^{\mathcal{E},n}_0)$, where $\mathcal{E}$ is a partition of the electrons into neutral clusters. The other cases follow analogously. Observe that
\begin{align}
    &P^\perp \tilde{H}_{N,M}(y^n) \tilde{S}(\psi^{\mathcal{E},n}_0) = \frac{1}{N! \lVert{S(\psi^{\mathcal{E},n}_0)}\rVert} \sum_{\pi} \sum_{i < l} P^\perp I_{\mathcal{E}(\pi)_i, \mathcal{E}(\pi)_l}(Y_{\mathcal{N}_i}^n, Y_{\mathcal{N}_l}^n) \psi^{\mathcal{E}(\pi),n}_0. 
\end{align}
Thus, by the previous argument using the expansion (\ref{E_0(yn)expansion3}), the estimates (\ref{interactionterms_ee_decay}) and (\ref{interactionterms_en_decay}), the CS inequality, and
(\ref{Hboresolvent_E0(yn)_estimate}), we obtain (\ref{FSresolventterm_estimate}).
\qed

%%%%%%%%%%%%%%%%%%%%%%%%%%%%%%%%%%%%%%%%%%%%%%%%%%%%%%%%%%%%%%%%%%%%%%%%%%%%
\appendix
%%%%%%%%%%%%%%%%%%%%%%%%%%%%%%%%%%%%%%%%%%%%%%%%%%%%%%%%%%%%%%%%%%%%%%%%%%%%

%%%%%%%%%%%%%%%%%%%%%%%%%%%%%%%%%%%%%%%%%%%%%%%%%%%%%%%%%%%%%%%%%%%%%%%%%%%%
\section{The time-independent Feshbach-Schur (FS) Map} \label{FSmap_subsection}
%%%%%%%%%%%%%%%%%%%%%%%%%%%%%%%%%%%%%%%%%%%%%%%%%%%%%%%%%%%%%%%%%%%%%%%%%%%%

In this subsection we introduce the FS map and formulate its main properties. See \cite{GD-IMS-BS} for further details, or \cite{GS}, Theorem 11.1 for a slightly different presentation. 

\begin{theorem} \label{FSmapTheorem}
Let $H$ be an on a Hilbert space, and $P$ and $P^\perp$ a pair of orthogonal projections such that $P + P^\perp = 1$. If
\begin{itemize}
    \item[1)] $\textrm{Ran } P \subset \mathcal{D}(H)$ and $\lVert{H P}\rVert < \infty$,
    \item[2)] the operator $(P^\perp H P^\perp - \lambda)$ is invertible,
\end{itemize}
then the operator 
\begin{equation}
    F_P(\lambda) := \left(PHP - P H P^\perp (P^\perp H P^\perp - \lambda)^{-1} P^\perp H P \right) \big|_{\textrm{Ran} P} \label{FPdef}
\end{equation}
is well-defined and has the following properties:
\begin{itemize}
    \item[a)] $\lambda \in \sigma(H)$ if and only if $\lambda \in \sigma(F_P(\lambda))$, 
    \item[b)] $H \psi = \lambda \psi$ if and only if $F_P(\lambda) \phi = \lambda \phi$, where $\phi = P \psi$ and $\psi = Q_P(\lambda) \phi$ and where 
    \begin{equation}
        Q_P(\lambda) := P - P^\perp (P^\perp H P^\perp - \lambda)^{-1} P^\perp H P. \label{QPdef}
    \end{equation}
    \item[c)] $\textrm{dim Null } (H - \lambda) = \textrm{dim Null } F_P(\lambda) $.
\end{itemize}
Finally, if $H$ is self-adjoint, then so is $F_P(\lambda)$. 
\end{theorem}

For a proof, see Theorem 11.1 in \cite{GS} and the proof there. In particular, we also make use of the following additional corollary, which follows from Theorem \ref{FSmapTheorem} part a). 

\begin{corollary} \label{FScorollary}
Let $H$ be self-adjoint and $\Lambda$ be an open subset of $\mathbb{C}$ such that $F_P(\lambda)$ is well-defined for all $\lambda \in \Lambda$, and let $\nu_i(\lambda)$, $\lambda \in \Lambda$, denote its eigenvalues counted with multiplicities, 
\begin{equation}
    \nu_1(\lambda) \leqslant \dots \leqslant \nu_m(\lambda).
\end{equation}
Then, a solution of the equation 
\begin{equation}
    \nu_i(\lambda) = \lambda, \lambda \in \Lambda \cap \mathbb{R},
\end{equation}
is the $i^{th}$ eigenvalue, $\lambda_i$, of $H$ and vice versa.
\end{corollary}

%%%%%%%%%%%%%%%%%%%%%%%%%%%%%%%%%%%%%%%%%%%%%%%%%%%%%%%%%%%%%%%%%%%%%%%%%%%%

\section*{Acknowledgments} The author is grateful to Ioannis Anapolitanos for suggesting the problem, suggesting edits to the manuscript to great improvement, and for many fruitful discussions and mentorship.

%%%%%%%%%%%%%%%%%%%%%%%%%%%%%%%%%%%%%%%%%%%%%%%%%%%%%%%%%%%%%%%%%%%%%%%%%%%%

\printbibliography

\end{document}